\documentclass[reprint,amsmath,amssymb,graphicx,longbibliography,showpacs]{revtex4-1}

\usepackage{amssymb}

\usepackage{amsfonts}
\usepackage{latexsym}

\usepackage{amsmath}
\usepackage{braket}
\usepackage{graphicx}

\usepackage{psfrag,pstricks}

\begin{document}
\title{Dispersive Qubit Measurement by Interferometry with Parametric Amplifiers}

\author{Sh. Barzanjeh$^{1}$}
\author{D. P. DiVincenzo$^{1,2}$}
\author{B.M. Terhal$^{1}$}
\affiliation{$^{1}$JARA-Institute for Quantum Information, RWTH Aachen University, 52056 Aachen, Germany \\
$^2$ Peter Gr\"{u}nberg Institute (PGI-2), Forschungszentrum J\"{u}lich, D-52425, J\"{u}lich, Germany.}
\date{\today}

\begin{abstract}
We perform a detailed analysis of how an amplified interferometer can be used to enhance the quality of a dispersive qubit measurement, such as one performed on a superconducting transmon qubit, using homodyne detection on an amplified microwave signal. Our modeling makes a realistic assessment of what is possible in current circuit-QED experiments; in particular, we take into account the frequency-dependence of the qubit-induced phase shift for short microwaves pulses. We compare the possible signal-to-noise ratios obtainable with (single-mode) SU(1,1) interferometers with the current coherent measurement and find a considerable reduction in measurement error probability in an experimentally-accessible range of parameters.
\end{abstract}

\pacs{42.50.Pq, 85.25.Hv, 85.25.Cp,03.67.-a}

\maketitle

\section{Introduction}

The realisation of quantum information processing in the laboratory requires quantum measurements of unprecedented speed and precision.  In particular, the measurements that will be necessary for achieving scalable fault tolerant quantum computing are understood in some detail, requiring 99+\% measurement fidelities, as well as the capability of repeating the same measurement very frequently within one coherence time ($T_2$) \cite{PhysRevA.80.052312}.  For the most part, these measurements are to be used as part of a quantum error correction scheme, to give an accurate diagnosis of corrections that are needed to maintain quantum coherence in the computation.  Fundamentally these ``syndrome'' measurements detect the parity (even or odd) of a collection of computational qubits; while special measurement schemes (so-called direct parity measurements) can be designed to access the particular multiqubit parities required \cite{LGB:2qubit_parity, divsol:parity, NG:stab, TBD:stoch_parity}, the quantum computation can be organised in such a way that single ancillary qubits hold the results of all necessary measurements.  In the present study we will focus only on the improvement of this basic single-qubit measurement.

We will also only focus here on the implementation of measurements in the setting of superconducting qubits within the paradigm of circuit quantum electrodynamics \cite{blais:meas} (cQED).  cQED techniques have contributed greatly to the quality of all aspects of quantum circuit implementation, measurement among them.  Single-qubit measurements were achieved before the advent of cQED -- a SQUID magnetometer strongly coupled to the qubit to be measured was biased to the edge of stability, so that it would switch to its normal state in a qubit-state dependent way.  This approach, while a great milestone in establishing the possibility of quantum computation in superconducting device systems, was unscalable, slow, very intrusive (i.e., far from quantum non-demolition (QND)), and also far from single-shot (fidelity $F$ far below 100\%).

With the advent of cQED, qubits with much longer coherence times have become available, and new, engineered forms of light-matter coupling have opened the possibility of higher quality measurements performed within the coherence times of the qubit.  The transmon qubit coupled to a high quality factor cavity realizes the Jaynes-Cummings model of atomic physics \cite{blais:meas}.  When the qubit transition frequency is off-resonance with respect to the cavity eigenmode frequency (``dispersive regime''), this cavity frequency is shifted by an amount dependent on the qubit state.  Probe radiation near this resonant frequency, transmitted or reflected from the cavity, acquires a phase shift $\varphi_+$ or $\varphi_-$ for qubit state $\ket{0}$ or $\ket{1}$.  The sensing of this phase shift accomplishes the quantum measurement, which will be QND so long as the probe radiation is weak enough that the conditions for the dispersive approximation for the Jaynes-Cummings model are met.  This condition will be an important constraint in the analysis that we give below; it is understood that ``high-power'' readout, involving the full nonlinearity of the Jaynes-Cummings model, can also give an effective (but non-QND) measurement \cite{Reed10}.

While it is not difficult to make the phase shift change large -- even $\varphi_+-\varphi_-=\pi$ is achievable -- the necessity for a weak probe means that the probe signal must be amplified before being mixed with a reference beam.  Fortunately, a reasonable amplifier in the necessary microwave band, the so called HEMT (``high electron mobility transistor'') has been available for low-temperature use, and has enabled qubits measurements near the single-shot regime \cite{Chow09}.  The HEMT remains essential in measurements up to the present, but it is far from ideal: its noise temperature around 10K prevents the achievement of genuinely high fidelity ($>90\%$) quantum measurements.

It was understood that, to go further, new types of superconducting devices would be needed to push the amplifier noise temperature into the desired millikelvin regime.  While the use of SQUIDs for low-noise amplifiers have been understood for a long time \cite{Clarke08}, the adoption of these devices in cQED setups, and the form of the amplifier used, has undergone steady evolution in recent years.  First, so-called ``bifurcation'' phenomena in modified qubits were used for initial amplification \cite{Sid04}.  From this work it was realised that further modifications of these devices would permit them to be used in parametric mode \cite{Kim}: the nonlinearity is used so that the device works as a linear but time-dependent circuit element.  Practical devices were made \cite{CBL07} and optimised in conjunction with extensive theoretical analysis \cite{Bergeal2010,Bergeal2010_2}.  These superconducting parametric amplifiers, operating very close to minimal noise temperatures, are now in use in many labs worldwide, with achievement of 99\% measurement fidelities now in sight.

Parametric devices have other functionalities besides amplification; they are also capable of producing squeezed radiation, which can be another tool in improving the noise performance of qubit systems.  Recently an experiment has been reported \cite{Murch13} in which squeezed radiation improves the coherence time of a transmon qubit.  Note that this involves having the probe radiation interacting with the parametric device {\em before} encountering the qubit-containing cavity -- when used as an amplifier, the parametric device comes {\em after} the probe has exited the cavity.

In this paper we explore the benefit gained from combining both, placing parametric devices {\em both before and after} the phase shifting element (qubit+cavity).  Such concepts were already explored in the pioneering work of Yurke and co-workers \cite{YMK}, who considered the possibility of such ``active'' interferometers, where the simple beamsplitters are replaced by active devices, both for optical and microwave systems.  This work defined the so-called ``SU(1,1)'' amplifiers, which we will describe and study in the present work \cite{foot1}.

To see how the SU(1,1) paradigm can be used to further improve qubit measurement, we will need to modify Yurke's approach to account for three aspects of the cQED setup: 1) phase shifts are not small,  2) probe radiation inside the cavity should be weak, 3) probe pulses $T_{pulse}$ should be of short duration, perhaps comparable to the inverse cavity linewidth $\kappa^{-1}$.  We will visit all these issues in the studies in this paper, showing improvements are indeed possible.

In \cite{YMK} two types of SU(1,1) interferometers were defined for which it was shown that they would give rise to a phase sensitivity $\Delta \varphi \sim \frac{1}{N}$ where $N$ is the total number of photons that pass through the interferometer: these interferometers are depicted in Fig.~\ref{fig:su11}(a) and (d). In these set-ups one measures the total number of outgoing photons $N_{out}$ so that 
\begin{equation}
\Delta \varphi \equiv \frac{\Delta N_{\rm out}}{|\partial N_{\rm out}/\partial \varphi|}.
\end{equation}
Here $(\Delta x)^2=\langle (x - \langle x \rangle)(x-\langle x \rangle)\rangle$ for an arbitrary operator or random variable $x$. 

Such scaling with $N$ is usually referred to as `reaching the Heisenberg limit' in contrast with the shot-noise limit $\Delta \varphi \sim \frac{1}{\sqrt{N}}$ which is reached by using a coherent state $\ket{\alpha}$ with average photon number $|\alpha|^2=N$ to determine the unknown phase shift, see e.g. \cite{LKD:rosetta} and \cite{GLM:metrology} and references therein. It is important to note that the enhanced phase-sensitivity is only reached for small phases $\varphi \approx 0$; in addition in the schemes in \cite{YMK} the input modes are taken to be in the vacuum state.

While the SU(1,1) interferometer is a way of obtaining a high sensitivity to an unknown phase shift, it does not immediately suit the experimental cQED setting for the following reasons. In the measurement chain for superconducting (transmon) qubits coupled to microwave cavities, the information-carrying signal is a microwave pulse which is amplified to a classical stochastic signal whose quadratures are recorded as classical voltages, see Section \ref{sec:mm}. This means that one does not measure the number of output photons of the interferometer, but rather the quadrature of one or both outgoing modes. By the linear optical transformation of the interferometer, any outgoing quadrature can be expressed as a linear function of the quadratures of the input modes. This means that the phase sensitivity of such quadratures is 0 when the input modes are prepared in the vacuum state and hence the quadrature signal carries no phase information~\cite{foot2}.

The simplest modification to this set-up is to provide the interferometer with a pulsed coherent microwave at one of its inputs, say, the mode $a_{1,in}$ in Fig.~(\ref{fig:su11}), which is what we will assume. We thus re-examine the phase sensitivity of the SU(1,1) interferometers under a homodyne measurement in Section \ref{sec:schemes} (see also \cite{li+:homodynesu11}). There are further features of the experimental set-up that we take into account; as mentioned above, the number of photons in the top arm of the interferometer should be bounded below a critical value in order for the measurement to be of non-demolition character. The number of photons coming out of the last PA or DPA in Fig.~\ref{fig:su11} should be sufficiently high so that further amplifications have a small effect. Thirdly, we wish the quantum measurement to be short: the finite time duration of the incoming pulse, $T_{\rm pulse}$ motivates the multi-frequency mode analysis in Section \ref{sec:mm} and \ref{sec:num_wig}.

\begin{figure}[ht]
\centering
\includegraphics[width=3in]{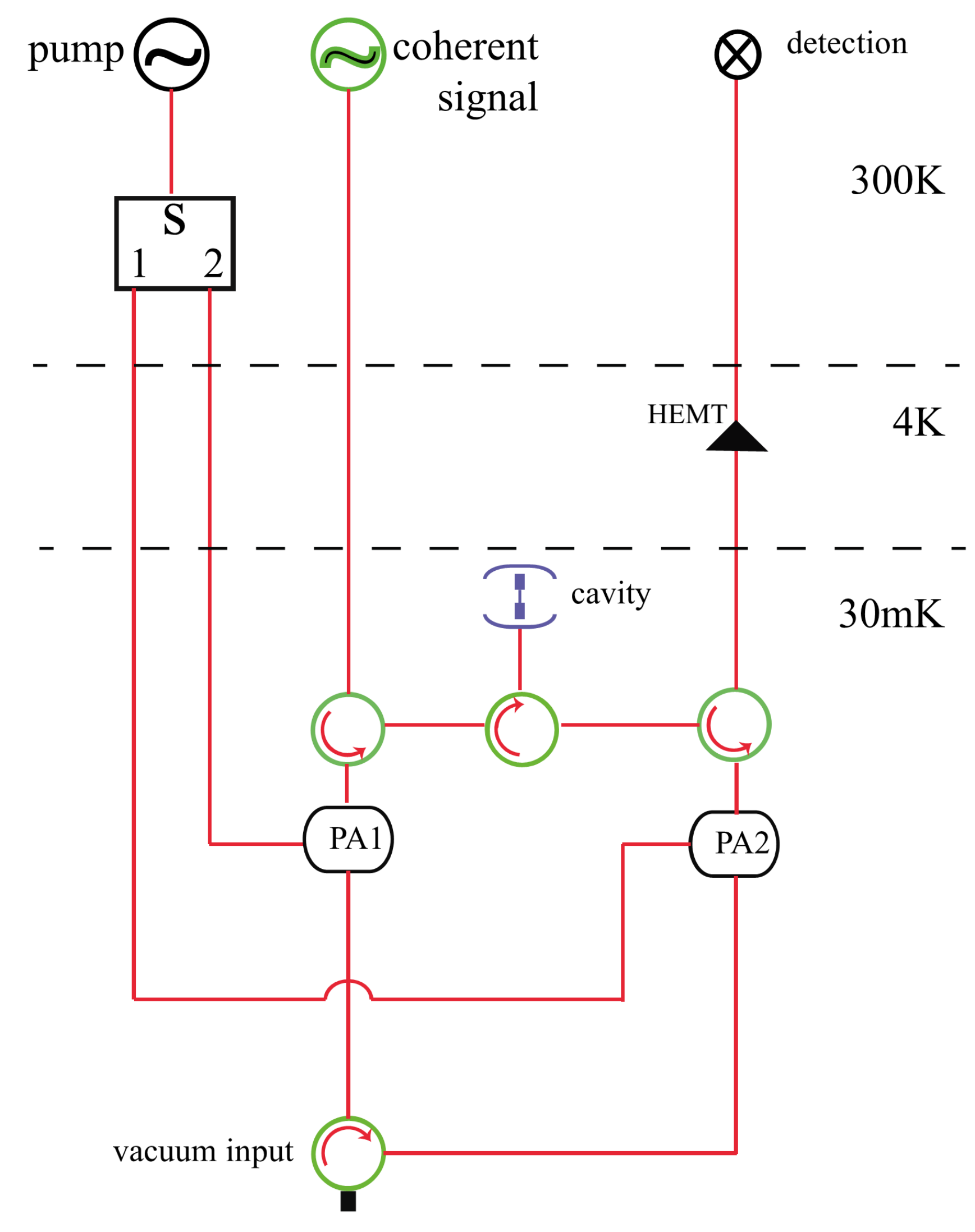}
\caption{Schematic of the experimental setup for implementing one version of the SU(1,1) interferometer (using non-degenerate parametric amplifiers) in present-day microwave components, which could be either lumped ``3D'' structures or on-chip integrated devices.  Many additional components of such an actual setup are omitted (e.g., attenuators, additional amplifiers, mixer local oscillators, data acquisition hardware); only the parts that are essential to our scheme are shown.  A pump is to be distributed by a power splitter to the non-degenerate two parametric devices (PA) (for example the ``Josephson parametric converter'' \cite{Bergeal2010}).  The PAs work in reflection, requiring circulators to separate input from output modes.  It is assumed that the cavity containing the qubit is probed in reflection.  The isolator/circulator at the bottom of the figure serves both to define the cold vacuum input to PA1 and to provide the necessary beam path from PA1 to PA2 for the reference mode of the SU(1,1) interferometer.  Relative phases of the pump beams, and of the two interferometer arms, must be precisely set, requiring careful choice of the propagation lengths along all these paths.  We show the HEMT amplifier (but not other amplifiers that would be involved in this setup) since it is necessary to consider whether the amplification provided by PA2 is large enough to overcome the non-ideal noise characteristics of the HEMT.}
\label{fig:expsu11}
\end{figure}

One reason to consider an SU(1,1) interferometer instead of a SU(2) Mach-Zehnder interferometer is that the attenuated microwave probe will have to undergo amplification anyhow in order to be detectable with current hardware; in this way the second amplifier in the interferometer does double duty (see however \cite{Kreysing} for a preliminary exploration of the Mach-Zehnder interferometer). It also means that the experimental set-up of the interferometer is not much more costly than the standard homodyne measurement in which typically only one Josephson-based amplifier is used, see Fig.~\ref{fig:su11}(c). We will find that the two-mode SU(1,1) interferometer which uses two non-degenerate parametric amplifiers, Fig.~\ref{fig:su11}(a), gives better results than a single-mode SU(1,1) interferometer, Fig.~\ref{fig:su11}(d), see Section \ref{sec:num_wig}: our proposed experimental setup is depicted in Fig.~\ref{fig:expsu11}. We note that in \cite{Teufel2009} the position of a nanomechanical oscillator, coupled to a microwave cavity, was measured in a `Mach-Zehner interferometric setup'. However, in that experimental set-up the signals from both arms of the interferometer are only recombined at room temperature allowing no entanglement between the arms of the interferometer. In our envisioned scheme the entire interferometer is realized at low temperature (e.g. $30$mK). In \cite{flurin+:su11} the authors used two non-degenerate Josephson parametric amplifiers (`Josephson mixers') to create a two-mode squeezed state which was subsequently analyzed by a second Josephson mixer: this set-up thus uses identical components as the SU(1,1) interferometer in Fig.~\ref{fig:su11} and shows that our proposal is experimentally feasible.\\

In the next section we will consider the four schemes in Fig.~\ref{fig:su11} with coherent state inputs and quadrature measurement on the mode $a_{1,out}$ or $a_{out}$ at the end. We assume the state of the qubit induces a phase shift 
\begin{eqnarray}
\varphi^{+}&=&+\varphi, \,\,\,\mbox{qubit state $\ket{0}$},\nonumber\\
\varphi^{-}&=&-\varphi, \,\,\,\mbox{qubit state $\ket{1}$},
\end{eqnarray} 
onto the passing probe (how it does this, is reviewed and analyzed in Section \ref{sec:mm}). Instead of focusing on the phase sensitivity, we derive expressions for the signal-to-noise ratio SNR, first assuming a simplified single-mode, single-frequency picture. This gives us insight in the gains that we can expect when we include the multi-mode nature of the input probe later on, in Sections \ref{sec:mm} and \ref{sec:num_wig}. We focus on the SNR as we do not expect $\varphi^{\pm}$ to be necessarily small, nor do we analyze the use of feedback in these schemes, but see Sec.~\ref{last} (Discussion). Let $x_{out}^{\pm}$ be any information-carrying quadrature and suppose that $\Delta x_{out}^+=\Delta x_{out}^-$ (we will restrict ourselves to such scenarios).  The signal-to-noise ratio is then given by
\begin{eqnarray}\label{eq:SNR}
{\rm SNR} \equiv \frac{|\langle x_{out}^+\rangle-\langle x_{out}^-\rangle|}{2\Delta x_{out}^{\pm}}.
\end{eqnarray}
This signal-to-noise ratio can be simply related to the probability of error of the quantum measurement, see Section \ref{sec:error}.  

\section{SU(1,1) Interferometers and Comparable Schemes}
\label{sec:schemes}

The action of an ideal phase-insensitive (also called `phase-preserving') non-degenerate parametric amplifier (PA), acting on four ports each of which is described by a continuum of modes labelled by frequency, is given \cite{YB:amp} by the following transformation
\begin{eqnarray}
\left(\begin{array}{c} b_{1,out}(\omega) \\ b_{2,out}^{\dagger}(2 \Omega-\omega) \end{array}\right)=S \left(\begin{array}{c} a_{1,in}(\omega) \\ a_{2,in}^{\dagger}(2 \Omega-\omega) \end{array} \right), \nonumber \\
S=\left(\begin{array}{cc} \cosh(r) & e^{i \theta} \sinh(r) \\ e^{-i\theta}\sinh(r) & \cosh(r) \end{array}\right),
\label{eq:pa}
\end{eqnarray}
where $\Omega$ is the frequency of the pump mode of the amplifier. The mode $a_{2,in}(\omega)$ functions as the `idler' mode and $a_{1,in}(\omega)$ as the `signal' mode. This transformation models a four-wave mixer in which 2 pump photons at frequency $\Omega$ are converted into one photon for mode 1 and one for mode 2, i.e. $2 \Omega=\omega_1+\omega_2$. Replacing $2\Omega$ by $\Omega$ in Eq.~(\ref{eq:pa}) would correspond to a three-wave mixer with $\Omega=\omega_1+\omega_2$.
 
Such a phase-insensitive non-degenerate PA amplifies both quadratures by the same amount, with a gain related to the real parameter $r$ by
\begin{equation}
G=\cosh^2 (r),
\end{equation} 
and this PA will always add a non-zero amount of noise \cite{caves:noise}. A good example of such an amplifier is the Josephson ring modulator \cite{Bergeal2010,Bergeal2010_2} used in the transmon qubit measurement in \cite{hatridge+:back}. The pump frequency $\Omega$ will be set at the carrier frequency $\omega_c$ of the microwave signal to be amplified.
In a phase-sensitive degenerate parametric amplifier (DPA) or squeezer the incoming modes $1,2$ are degenerate and the amplifier thus enacts the following idealized transformation on a single frequency mode, given the pump frequency $\Omega$: 
\begin{equation}
b_{out}(\omega)= \cosh(r)a_{in}(\omega)+ e^{i \theta} \sinh(r) a_{in}^{\dagger}(2\Omega-\omega).
\label{eq:dpa}
\end{equation}
Such an amplifier will squeeze the outgoing quadratures and will add a corresponding quadrature-dependent amount of noise. Very good phase-sensitive Josephson parametric amplifiers have been developed in \cite{cast2008, thesis:cast}; for such amplifiers the dominant source of noise at the output of the amplifier is the quantum fluctuations of the ingoing weak signal. For both phase-sensitive and phase-insensitive amplifiers we assume that the phase $\theta$ and gain $G$ are independent of frequency $\omega$. This approximation is warranted for the usual operating conditions of the current microwave devices which have sufficiently large bandwidth $\times$ gain characteristics.

\begin{figure}[htb]
\centering
\includegraphics[width=3.4in]{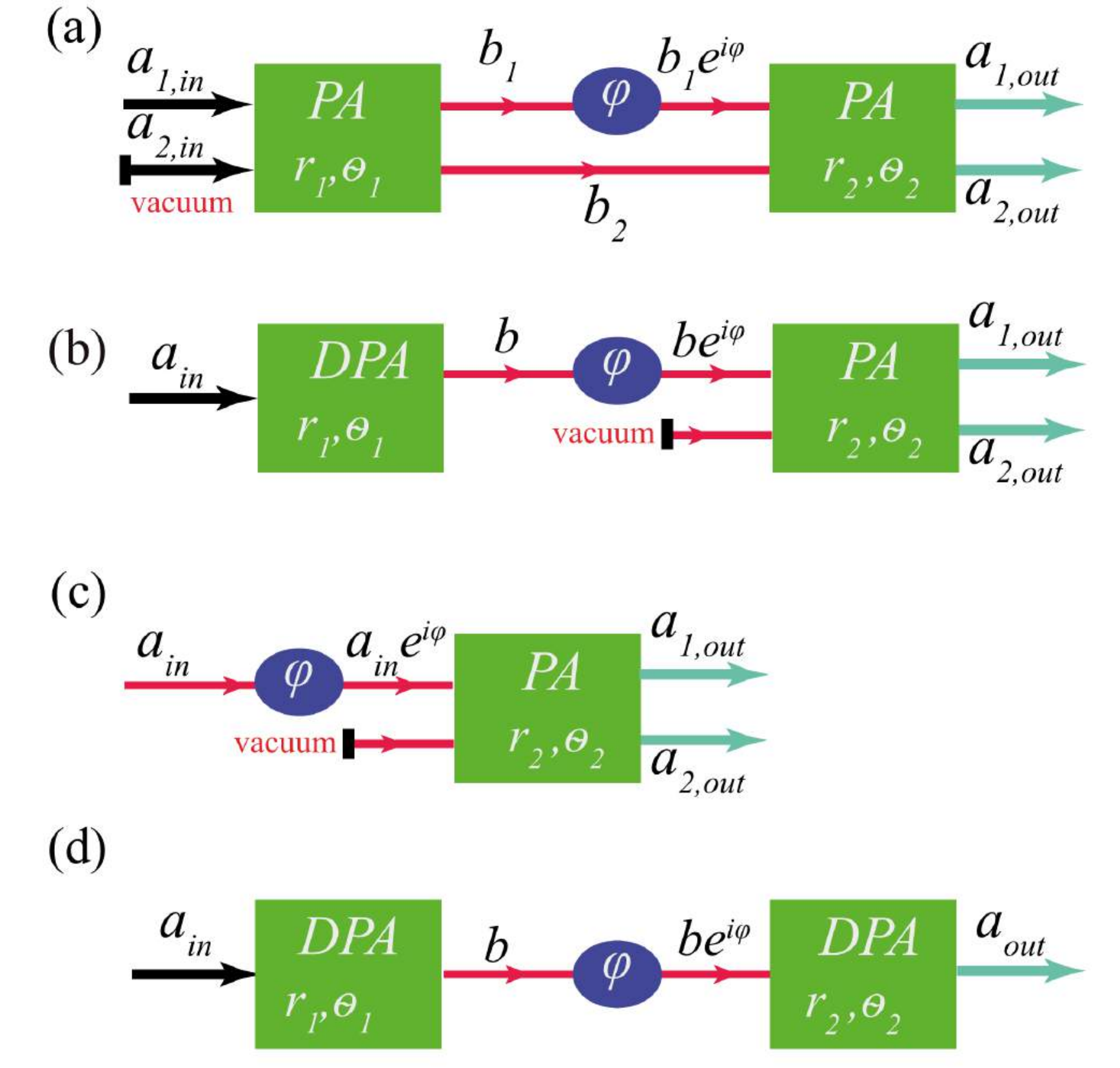}
\caption{Possible scenarios of a dispersive qubit measurement in a circuit-QED setting where $\varphi$ is the qubit-state dependent phase shift. The pump beams of the parametric (PA) and degenerate parametric amplifiers (DPA) are not explicitly depicted. (a) The (two-mode) SU(1,1) interferometer. When both input modes are prepared in the vacuum state and the parametric amplifiers are chosen such that $r=r_1=r_2$, $\theta_1-\theta_2=\pi$, the total number of photons at output $N_{out}=N_{1,out}+N_{2,out}$ is a sensitive probe for the phase $\varphi$, i.e. $(\Delta \varphi)^2=\frac{1}{\sinh^2(r)}$ at $\varphi=0$ \cite{YMK}. In our scenario we consider a coherent pulse in mode $a_{1,in}$ and a homodyne measurement is done on the outgoing mode $a_{1,out}$. (b) The squeeze scenario in which a coherent input pulse is first squeezed by a degenerate parametric amplifier before picking up a to-be-determined phase-shift. A phase-insensitive amplifier subsequently amplifies the signal so that homodyne measurement is possible. (c) The standard coherent dispersive measurement scenario in which a coherent pulse, after having picked up a phase-shift at the cavity, is amplified. (d) The single-mode SU(1,1) interferometer in which two degenerate parametric amplifiers sharing the same pump are used. The relative difference of the phases $\theta_1$ and $\theta_2$ is determined by the pump beam.}
\label{fig:su11}
\end{figure}

\begin{figure}[ht]
\centering
\includegraphics[width=3.5in]{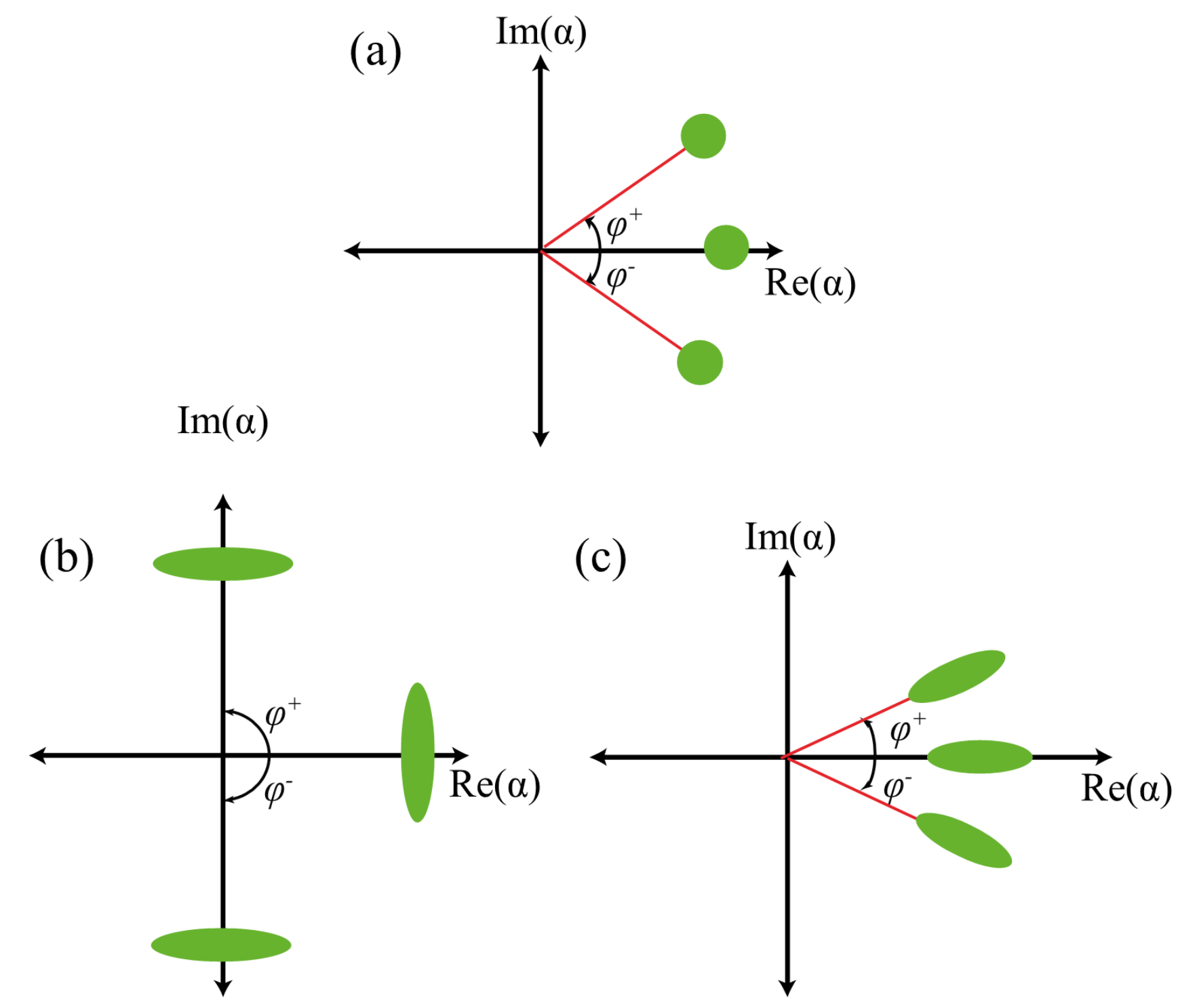}
\caption{Phase-space sketches. In all three figures we start with a coherent or squeezed state with $\langle p \rangle={\rm Im}(\alpha)=0$ and the state picks up a phase-shift $\varphi^{\pm}$. (a) A coherent state picks up one of two phases depending on the state of a qubit, $\ket{\alpha} \rightarrow \ket{\alpha e^{\pm i\varphi}}$. Values $|\varphi^+-\varphi^-| > \pi$ do not give rise to greater distinguishability when measuring ${\rm Im}(\alpha)$, but as we will find in this paper modeling the input as a multi-mode signal and constraining the number of photons in the cavity will lead to a better SNR for larger phase shifts. (b) and (c) The optimal direction of squeezing depends on the phase shift which is $\pi/2$ in (b) and close to $0$ in (c). It is clearly not possible to minimize the noise for both $\pm$ signals when $\varphi^{\pm}$ is not close to the points $0$ or $\pi/2$.}
\label{fig:ps_sketch}
\end{figure}

In the remainder of this section we will give closed-form expressions for SNR for the setups of Fig.~\ref{fig:ps_sketch} in the simple one-frequency approximation that is standard in quantum optics; the next section will give the full multi-mode analysis.  Thus, for the non-degenerate parametric amplifier, we take $\Omega=\omega$ and write Eq.~(\ref{eq:pa}) without frequency arguments:
\begin{equation}
\left(\begin{array}{c} b_{1,out} \\ b_{2,out}^{\dagger} \end{array}\right)=\left(\begin{array}{cc} \cosh(r) & e^{i \theta} \sinh(r) \\ e^{-i\theta}\sinh(r) & \cosh(r) \end{array}\right)\left(\begin{array}{c} a_{1,in} \\ a_{2,in}^{\dagger}\end{array}\right) .
\label{eq:pa4}
\end{equation}
In other words, this relation is considered to be one involving just four modes, rather than four continua of modes.  The degenerate parametric amplifier relation Eq.~(\ref{eq:dpa}) is likewise simplified to one involving only two discrete modes:
\begin{equation}
b_{out}= \cosh(r)a_{in}+ e^{i \theta} \sinh(r) a_{in}^{\dagger}.
\label{eq:dpa2}
\end{equation}

We first consider the SNR of the current measurement schemes through which a qubit is measured, see e.g. \cite{hatridge+:back}, schematically depicted in Fig.~\ref{fig:su11}(c). A coherent microwave pulse picks up a phase shift at the cavity, see the sketch in Fig.~\ref{fig:ps_sketch}(a), after which the signal is amplified by a single phase-insensitive Josephson parametric amplifier whose mode transformation is given in Eq.~(\ref{eq:pa4}). We thus assume an input state $\ket{\alpha}$ (fixing $\mbox{Im}(\alpha)=0$) in mode $1$ so that the outgoing $p$-quadrature $p_{1,out}$ contains the maximal amount of information. We use the quadrature convention $p= -i(a-a^{\dagger})/\sqrt{2}$, implying $(\Delta p)^2=\frac{1}{2}$ for any coherent state. In this scenario the SNR defined in Eq.~(\ref{eq:SNR}) can be calculated as 
\begin{equation}
{\rm SNR}_{\rm coherent+PA}=\frac{2\sqrt{n_{in}} |\sin(\varphi)|}{\sqrt{2 A_N+1}},
\label{eq:SNRcoh}
\end{equation}
with added noise number $A_N=\frac{1}{2}(1-G^{-1})$ (see also Eq.~(\ref{eq:ampnoise})) and $n_{in}=|\alpha|^2$. The Haus-Caves theorem \cite{caves:noise} states that for a non-ideal phase-insensitive parametric amplifier $A_N \geq \frac{1}{2}|1-G^{-1}|$ where equality is achieved for a vacuum state at the idler port. We see that the SNR corresponds to shot-noise behavior, i.e. ${\rm SNR} \sim \sqrt{n_{in}}$. The expression does not depend on the phase $\theta_1$ of the amplifier as the amplifier adds noise to each quadrature by the same amount. 

Alternatively, one can use an ideal degenerate parametric amplifier at the output for a coherent input signal. We obtain
\begin{eqnarray}
\lefteqn{\langle p_{out}\rangle=\sqrt{2 n_{in}}\left(\cosh(r) \sin(\varphi)+\sinh(r) \sin(\theta-\varphi)\right),} \nonumber \\
 & & (\Delta p_{out})^2=\frac{1}{2}\left(\cosh(2r)-\cos(\theta)\sinh(2r)\right),
\label{eq:cohDPA}
\end{eqnarray}
and a corresponding ${\rm SNR}_{\rm coherent+DPA}$. At $\theta=\pi$, one obtains the expected ${\rm SNR}_{\rm coherent+DPA}=2 \sqrt{n_{in}}|\sin(\varphi)|$, showing that the DPA does not add any additional noise. Note that for $\theta=0$, both the noise and the signal are vanishingly small.\\

Next we consider the two-mode SU(1,1) interferometer in Fig.~\ref{fig:su11}(a). We can obtain the composite mode transformation of the two amplifiers and the phase shift modeled by the matrix $\left(\begin{array}{cc} e^{\pm i \varphi} & 0 \\ 0 & 1\end{array} \right)$. Again we assume $\ket{\alpha}$ (with ${\rm Im}(\alpha)=0$) in mode $1$ and obtain general expressions for $(\Delta p_{1,out})^2$ and $\langle p_{1,out} \rangle$ as 
\begin{eqnarray}
\langle p_{1,out} \rangle=\sqrt{2 n_{in}} \left[\cosh(r_1) \cosh(r_2) \sin(\varphi)+\right.\nonumber\\
\left.\sinh(r_1) \sinh(r_2) \sin(\theta_2-\theta_1)\right], \nonumber
\end{eqnarray}
and
\begin{eqnarray}
(\Delta p_{1,out})^2=\frac{1}{2}\left[\cosh(2r_1) \cosh(2r_2)+ \right.\nonumber \\
\left.\cos(\theta_1-\theta_2+\varphi) \sinh(2r_1) \sinh(2r_2)\right].\nonumber
\label{eq:signal_noise}
\end{eqnarray}
One can observe that the signal $|\langle p_{1,out}^+\rangle-\langle p_{1,out}^-\rangle|$ does not depend $\theta_1-\theta_2$. The noise $\Delta p_{1,out}^{\pm}$ is clearly minimized when $\theta_1-\theta_2+\varphi^{\pm}=\pi$, but this condition cannot be satisfied for both measurement outcomes simultaneously when $|\varphi^{\pm}| > 0$. An optimal choice is to take $\theta_1-\theta_2=\pi, \theta_1=0$ for $|\varphi^{\pm}| \leq \frac{\pi}{2}$ and $\theta_1=\theta_2=0$ for $|\varphi^{\pm}| > \frac{\pi}{2}$: in both settings $\Delta p_{1,out}^+=\Delta p_{1,out}^-$, see Fig.~\ref{compareSNR} in Section \ref{sec:num_wig}. The expression for the SNR for the choice $\theta_1-\theta_2=\pi$ equals
\begin{eqnarray}
\lefteqn{{\rm SNR}_{\rm SU(1,1)+PA}=2\sqrt{n_{in}} |\sin(\varphi)|\times} \nonumber \\
& & \frac{1}{\sqrt{(2A_N^1+1)(2A_N^2+1)-8 \cos(\varphi)\sqrt{A_N^1 A_N^2}}},
\label{eq:SNRsu11_1}
\end{eqnarray}
where $A_N^1$ ($A_N^2$) are the noise quanta added by the first (the second) amplifier. If the first amplifier has $G_1=1$ (no amplification), we have $A_N^1=0$ and recover ${\rm SNR}_{\rm coherent+PA}$. For a small phase around $\varphi^{\pm} \approx 0$ and $A_N^1 \approx A_N^2$ one has ${\rm SNR}_{\rm SU(1,1)+PA}=\frac{2 \sqrt{n_{in}} |\sin(\varphi)|}{2A_N-1}$ where the noise vanishes in the limit of large gain, $A_N \rightarrow \frac{1}{2}$. Note that for $|\varphi^{\pm}| \approx \pi/2$, for which the signal is maximal, the noise in the denominator of Eq.~(\ref{eq:SNRsu11_1}) does not get suppressed: when $A_N^1=A_N^2=\frac{1}{2}$, the SNR of the interferometer is $1/\sqrt{2}$ worse as compared to the coherent state expression in Eq.~(\ref{eq:SNRcoh}), due to the added noise of the first amplifier.

It is of interest to compare this two-mode SU(1,1) interferometric setup with other uses of two (Josephson) parametric amplifiers depicted in Fig.~\ref{fig:su11}(b) and (d). In scenario (b) the first degenerate PA squeezes the incoming signal before it interacts with the qubit according to the mode transformation in Eq.~(\ref{eq:dpa}). The signal emerging from the cavity is then amplified by a non-degenerate PA after which a homodyne measurement is done. The difference between this {\em squeeze} scenario and the SU(1,1) interferometer in Fig.~\ref{fig:su11}(a) is that the probe state is not entangled between two modes. This means that the SNR will not depend on the relative phase $\theta_1-\theta_2$. One does expect an improvement in SNR as compared to ${\rm SNR}_{\rm coherent+PA}$ since pre-squeezing can reduce the noise in the information-carrying quadrature of the outgoing signal. One has 
\begin{eqnarray}
\langle p_{1,out} \rangle=\sqrt{2 n_{in}} \cosh(r_2) \left[\cosh(r_1)\sin(\varphi)+\right.\nonumber \\
\left.\sinh(r_1)\sin(\theta_1+\varphi)\right],\nonumber
\end{eqnarray}
and
\begin{eqnarray}
\lefteqn{(\Delta p_{1,out})^2=\frac{1}{2}\left[\sinh^2(r_2)+\right.}\nonumber \\
& & \left.\cosh^2(r_2)\left(\cosh(2r_1)-\sinh(2r_1)\cos(2\varphi+\theta_1)\right)\right].\nonumber
\end{eqnarray}
For $r_1=0$ we again obtain the coherent SNR. The optimal direction of squeezing which is determined by $\theta_1$ depends on how large the phase shift $\varphi$ is, see the sketches in Fig.~\ref{fig:ps_sketch}(b) and (c).  For very small $\varphi \approx 0$, $\theta_1$ should be chosen to be 0 to minimize $(\Delta p_{1,out})^2$. On the other hand, for $\varphi=\pm \pi/2$, the noise is minimized for both $\Delta p_{1,out}^{\pm}$ for $\theta_1=\pi$ and the signals $\langle p_{1,out}^{\pm}\rangle$ differ by the maximal amount. For $\varphi$ away from these points, the optimal noise-minimizing squeezing direction is different for $\pm \varphi$. If we require that $\Delta p_{1,out}^+=\Delta p_{1,out}^-$ we can choose $\theta_1=0$ for $|\varphi^{\pm}| \leq \frac{\pi}{4}$ and $\theta_1=\pi$ for $|\varphi^{\pm}| > \frac{\pi}{4}$ so that $\cos(2 \varphi+\theta_1) \geq 0$. Choosing $\theta_1=0$, we obtain
\begin{eqnarray}
{\rm SNR}_{\rm squeeze}=\frac{2 \sqrt{n_{in}}\left(1+\sqrt{1-G_1^{-1}}\right) |\sin(\varphi)|}{\sqrt{G_1^{-1}(2A_N^2-1)+2-2\sqrt{1-G_1^{-1}}\cos(2\varphi)}}, \nonumber
\end{eqnarray}
where no squeezing, so a coherent state input, corresponds to the case $G_1=1$, giving the coherent SNR. Clearly, the SNR can be enhanced in this scenario for sufficiently large $G_1$, but this gain $G_1$ is limited as we need to bound the number of photons interacting with the qubit in the cavity and thus the noise contribution proportional to $G_1^{-1}$ may not be negligible. We will not analyse this `squeeze' scenario in more detail as more favourable SNRs can probably be obtained by the use of DPAs in an interferometric set-up.

Hence in our last scenario, that of the single-mode SU(1,1) interferometer \cite{YMK}, both parametric amplifiers are degenerate, see Fig.~\ref{fig:su11}(d). In the regime $\varphi \approx 0$, Ref.~\cite{YMK} has shown that this interferometer can also reach the Heisenberg limit if photon-number measurements are assumed. Choosing $\theta_1=0$ and $\theta_2=\pi$ as in \cite{YMK} one can obtain
\begin{eqnarray}
\lefteqn{\langle p_{out}\rangle=\sqrt{2 n_{in}} e^{r_1+r_2} \sin(\varphi)}, \nonumber \\
& & (\Delta p_{out})^2=\frac{1}{2}\left[e^{2r_2}(\cosh(2r_1)-\cos(2\varphi)\sinh(2r_1))\right],\nonumber
\end{eqnarray}
giving 
\begin{equation}
{\rm SNR}_{\rm SU(1,1)+DPA}=\frac{2\sqrt{n_{in}} |\sin(\varphi)|}{\sqrt{\frac{1}{2}(1-\cos(2 \varphi))+\frac{1}{2}(1+\cos(2\varphi))e^{-4 r_1}}}.
\label{eq:SNRsu11_2}
\end{equation}
We note that this SNR does not depend on the gain of the second amplifier (we assume that it is an ideal amplifier, adding no noise), but the second amplifier will be needed to process the signal in any case. For small $\varphi^{\pm} \approx 0$, the noise vanishes as $\exp(-4 r_1)$ corresponding to the Heisenberg limit. When $\varphi=\pm \frac{\pi}{2}$, the SNR equals $2 \sqrt{n_{in}}$ which is identical to the ${\rm SNR}_{\rm coherent +DPA}$. Comparing it with ${\rm SNR}_{\rm coherent+PA}$ we see that the coherent SNR is worse by a factor $1/\sqrt{2}$ due to the added noise.

It is clear that the noise is reduced as compared to a coherent measurement when $\cos(2\varphi) > 0$, that is, for small angles $0 \leq |\varphi^{\pm}| \leq \frac{\pi}{2}$ or relatively large angles $\frac{5 \pi}{4} \geq |\varphi^{\pm}| \geq \frac{3 \pi}{4}$. We note that for $G_1 \rightarrow \infty$, the expression for ${\rm SNR}_{\rm SU(1,1)+DPA}$ coincides with ${\rm SNR}_{\rm squeeze}$.

In Section \ref{sec:num_wig} we present numerical values for these various signal-to-noise ratios within a full multi-mode analysis and show the qualitative improvement of the SU(1,1) interferometer using further details of the modelling of the qubit measurement.

We note that in all these scenarios we have, as stated above, assumed that $\varphi^{\pm}=\pm \varphi$ and taken the $p$-quadrature of the outgoing signal. For the two-mode SU(1,1) interferometer, one can show that the expression of a different outgoing 
($\delta$-rotated) quadrature is identical to the expression for the $p$-quadrature when we phase-shift $\varphi^{\pm}=\delta \pm \varphi$, and change the phase of the last parametric amplifier by $\delta$. Choosing the $p$-quadrature when $\varphi^{\pm}=\pm \varphi$ is intuitively optimal, see Fig.~\ref{fig:ps_sketch}, but we also have verified numerically that this is optimal for the schemes that we consider in Section \ref{sec:num_wig}.

\section{Description of the Circuit-QED Measurement Chain}
\label{sec:mm}

In this section we will review several details of the description of a dispersive quantum measurement of a qubit. We consider a superconducting qubit with resonance frequency $\omega_q$  such as the transmon qubit which is capacitively coupled to a 2D or 3D microwave cavity. We assume that a particular cavity mode $a$ with resonance frequency $\omega_r$ couples most strongly to the qubit and neglect the interactions of the qubit with other cavity modes, as well as the coupling to all higher-energy levels beyond the states $\ket{0}$ and $\ket{1}$. The interaction between qubit and cavity mode is then approximately described by the Jaynes-Cummings Hamiltonian $H_{JC}=-\frac{\hbar\omega_q}{2} Z+\hbar \omega_r a^{\dagger}a+\hbar g (\sigma_- a^{\dagger}+\sigma_+ a)$ where $ Z $ is the Pauli matrix. The eigenstates of the Jaynes-Cummings model are entangled states between transmon-qubit and cavity mode, but in the dispersive regime when $\frac{g}{\Delta} \ll 1$ ($\Delta=\omega_q-\omega_r$ is the detuning), one may do a perturbative expansion in $\frac{g}{\Delta}$ and derive an effective Hamiltonian via a Schrieffer-Wolff transformation (see e.g. \cite{richer:thesis}, typo corrected here) to obtain
\begin{eqnarray}
{1\over\hbar}\lefteqn{H_{\rm eff}=(\omega_r+\frac{5g^4}{\Delta^3}-\chi Z)a^\dagger a-\frac{1}{2} (\omega_q+\chi)Z}\nonumber \\ 
& & +\frac{5g^4}{3\Delta^3}Z (a^{\dagger}a)^2+O\left(\frac{g^6}{\Delta^5}\right), \nonumber \\
& & \chi=\frac{g^2}{\Delta}+\frac{5g^4}{6\Delta^3}+O\left(\frac{g^6}{\Delta^5}\right)
\label{eq:heff}
\end{eqnarray}
Such expansion is warranted for $\frac{2g \sqrt{\overline{n}+1}}{\Delta} \ll 1$ where $\overline{n}$ is the average number of photons in the cavity. We note that due to the multi-level nature of the transmon qubit, the dispersive shift is more accurately given by $\chi\simeq - E_c g^2/(\Delta(\Delta-E_c))$~\cite{koch+:purcell}, where $E_c$ is the charging energy of the Cooper pair box. 

It is essential for our analysis that we remain within the regime of validity of this picture, which breaks down when the number of photons in the cavity is beyond a critical photon number (as estimated in the two-level approximation for the transmon quit) $\overline{n} > n_{crit}=\Delta^2/(4g^2)$; in this regime the eigenstates of the Hamiltonian are entangled `atom' and cavity field states. If we wish to use the interaction with the cavity mode to perform a quantum measurement, such measurement will thus change the state of the qubit and will cease to be of non-demolition character. We would like to be considerably into this regime $\overline{n} < n_{crit}$ so that neglecting the nonlinear term $\propto Z(a^{\dagger}a)^2$ is also warranted: it has been shown in \cite{BGB:nonlin_disp} that such nonlinear coupling can lead to a reduction in SNR.

The effective Hamiltonian shows that the resonant frequency of the cavity is shifted depending on the state $\ket{0}$ (+) or $\ket{1}$ ($-$) of the qubit, i.e. its frequency 
\begin{equation}
\omega_r \rightarrow \tilde{\omega}_r=\omega_r \mp \chi+O\left(\frac{g^4}{\Delta^3}\right).
\label{eq:defomega}
\end{equation}
Detecting this frequency shift thus amounts to a dispersive, non-demolition, measurement of the qubit  state in the $\ket{0},\ket{1}$ basis.


We imagine that a microwave transmission line is capacitively coupled to the cavity on one side only, i.e. radiation enters and leaves the cavity through the same port or we use the cavity `in reflection' (see Fig.~\ref{fig:expsu11}). This can be achieved by having outgoing transmission line couple asymmetrically to the cavity, see e.g. \cite{hatridge+:back} where $\kappa_{in} \ll \kappa_{out}$ determine the decay rates on both sides, or having a tunable coupler to the cavity \cite{sete+eyob} or simply having one ingoing transmission line. The cavity can be a 1D stripline cavity \cite{blais:meas} or a 3D cavity \cite{paik:3Dcoherence}. The strength with which the cavity mode $a$ interacts with the continuum of modes in the one-dimensional transmission line will determine the cavity decay rate $\kappa$. We will neglect other sources of cavity decay in our modelling. Furthermore, we neglect qubit decoherence during the measurement because the transmon qubit coherence time are $O(10)\mu$sec or more \cite{paik:3Dcoherence,chang_IBM:highcoh}, much longer than the measurement times that we will consider. Table \ref{table:parameters} shows the experimental range of values of the relevant parameters.

The linear weak coupling of a single cavity mode to a continuum of travelling modes for a one-dimensional transmission line is modelled using input-output theory \cite{book:walls_milburn, clerk+:rmp, book:gardiner_zoller}. Neglecting the nonlinear terms in $H_{\rm eff}$, the cavity acts as a linear optical device whose effect can be described on a set of frequency-labelled ingoing and outgoing modes, see the background details in Appendix \ref{sec:qmc_details}. When one eliminates the cavity field one obtains a direct relation between an input mode $b_{in}(\omega)$ at frequency $\omega$ and an output mode $b_{out}(\omega)$ (defined as the Fourier transform of the Heisenberg operator $b_{in}(t)$ resp. $b_{out}(t)$, see Appendix \ref{sec:qmc_details}), viz. 
\begin{equation}\label{eq:phase_out}
b_{out}(\omega)=\frac{\kappa/2+i(\omega-\tilde{\omega}_r)}{\kappa/2-i(\omega-\tilde{\omega}_r)}b_{in}(\omega)=e^{i\varphi^{\pm}(\omega-\omega_{r})}b_{in}(\omega),
\end{equation}
where $\tilde{\omega}_r=\omega_r\pm\chi$. The presence of the qubit in the cavity thus induces a state-dependent phase shift on the outgoing signal $b_{out}(\omega)$ given by  \cite{blais:meas}
\begin{equation}
\varphi^{\pm}(\omega-\omega_r)=2 \,\mathrm{arctan}\Big[\frac{2(\omega-\omega_r)}{\kappa}\pm \frac{2 \chi}{\kappa} \Big].
\label{eq:phaseshift}
\end{equation}

If one drives the cavity at resonance $\omega_r$, the phase-shifts equal $\varphi^{\pm}=\pm 2 \arctan(\frac{2 \chi}{\kappa})$, symmetric around $0$. Maximal distinguishability with a quadrature measurement would be achieved with 
$|\varphi^+-\varphi^-|=\pi$ difference corresponding to $2 \chi=\frac{2 g^2}{\Delta}=\kappa$. We have also seen in the single-frequency mode SNR expressions,  Eqs.~(\ref{eq:SNRsu11_1},\ref{eq:SNRsu11_2}), that for such an optimal phase shift, the benefits of interferometers and squeezings are negligible. 

However, two aspects of the realization of this measurement alter this picture. First of all, for reasonably short pulses --- and it is the goal to have a short measurement time --- one needs to take into account the frequency dependence of the phase shift $\varphi^{\pm}(\omega-\omega_r)$. Secondly, we need to work under the condition that the number of photons in the cavity at any given time $\overline{n}(t) \ll n_{crit}$. Let us consider these issues in more detail.

The expression for the cavity field $a(\omega)$ (defined as the Fourier transform of the Heisenberg operator $a(t)$) equals
\begin{equation}
a(\omega)=\frac{\sqrt{\kappa}}{\frac{\kappa}{2}-i (\omega-\omega_r)} b_{in}(\omega). \nonumber
\end{equation}
Hence the expected number of photons in the cavity $\overline{n}(t)$ as a function of time is given by 
\begin{eqnarray}
\lefteqn{\overline{n}(t)=\langle a^{\dagger}(t) a(t) \rangle=\frac{1}{2\pi}\int_{\infty}^{\infty} d\omega \int_{-\infty}^{\infty} d\omega'\, e^{i(\omega -\omega')t}} \nonumber \\
& & \frac{\kappa\langle b_{in}(\omega)^{\dagger} b_{in}(\omega')\rangle}{(\kappa/2-i(\omega-\tilde{\omega}_r))(\kappa/2+i(\omega'-\tilde{\omega}_r))}.
\label{eq:ncav}
\end{eqnarray}

For a simple plane-wave coherent state travelling towards the cavity with wavenumber $k_c > 0$ and frequency $\omega_c=v k_c$, we have $\langle b_{in}(\omega)^{\dagger} b_{in}(\omega') \rangle=\delta(\omega-\omega_c) \delta(\omega'-\omega_c) 2 \pi F_t$ where $F_t$ is the photon-flux per unit time, see Appendix \ref{sec:qmc_details}.  For such a plane-wave input, one has
\begin{equation}
\overline{n}(t)=\frac{\kappa F_t}{\frac{\kappa^2}{4}+(\omega_c-\tilde{\omega}_r)^2} \underset{\omega_c=\omega_r}{\rightarrow} \frac{\kappa F_t}{\frac{\kappa^2}{4}+\chi^2}.
\label{eq:ncav}
\end{equation}
From this Lorentzian profile of $\overline{n}(t)$, it is clear that the larger the value of $\frac{2 \chi}{\kappa}$, the further one is removed from the resonance at $\omega=\tilde{\omega}_r$, the lower the number of photons in the cavity at a given point in time.  Given a fixed upper value for the photon number in the cavity $\bar n$, the number of input photons $n_{in}$ (proportional to the flux $F_t$ in Eq. (\ref{eq:ncav}) ) is an increasing function of $2 \chi/\kappa$, i.e., as the system is taken further from resonance.  Thus, the optimal value $2 \chi/\kappa$ for the SNR expressions in Sec. II can, and does, exceed the value $2 \chi/\kappa=1$ for which the phase shift per photon is optimal.  Each photon is less informative, but we can safely send more of them through the system.


Another effect, as we will see numerically in Section \ref{sec:num_wig}, is the effect of dispersion due to the finite pulse time. Any incoming microwave pulse signal of finite duration $T_{\rm pulse}$ has a non-zero frequency spread $W$. We choose such a pulse to have its center frequency at the bare resonance frequency, i.e. $\omega_c=\omega_r$, such that (see Appendix \ref{sec:qmc_details})
\begin{equation}
\alpha(\omega)=\frac{\alpha_0 \,e^{-(\omega-\omega_c)^2/W^2}}{(2\pi)^{1/4}\sqrt{W/2}}, \langle b_{in}^{\dagger}(\omega) b_{in}(\omega')\rangle=\alpha^*(\omega) \alpha(\omega'), 
\label{eq:gauss}
\end{equation}
where the total number of photons in the input pulse is 
\begin{equation}
n_{\rm pulse}= \int d\omega \; |\alpha(\omega)|^2
\label{defpul} 
\end{equation}
while $|\alpha(\omega)|^2$ is the photon flux per unit angular frequency at frequency $\omega$ (thus in units of seconds).
If we consider the intensity $|\alpha(\omega)|^2$ of this pulse per unit angular frequency, we see that this is a Gaussian with standard deviation $W/2$. If we Fourier transform $\alpha(\omega)$ to $\alpha(t)$ and consider the intensity of the pulse per unit time $|\alpha(t)|^2$, we note that it has a standard deviation of $1/W$ and thus we can take $T_{\rm pulse}=2/W$ as a measure of the time duration of the pulse.

\begin{figure}[ht]
\centering
\includegraphics[width=2.5in]{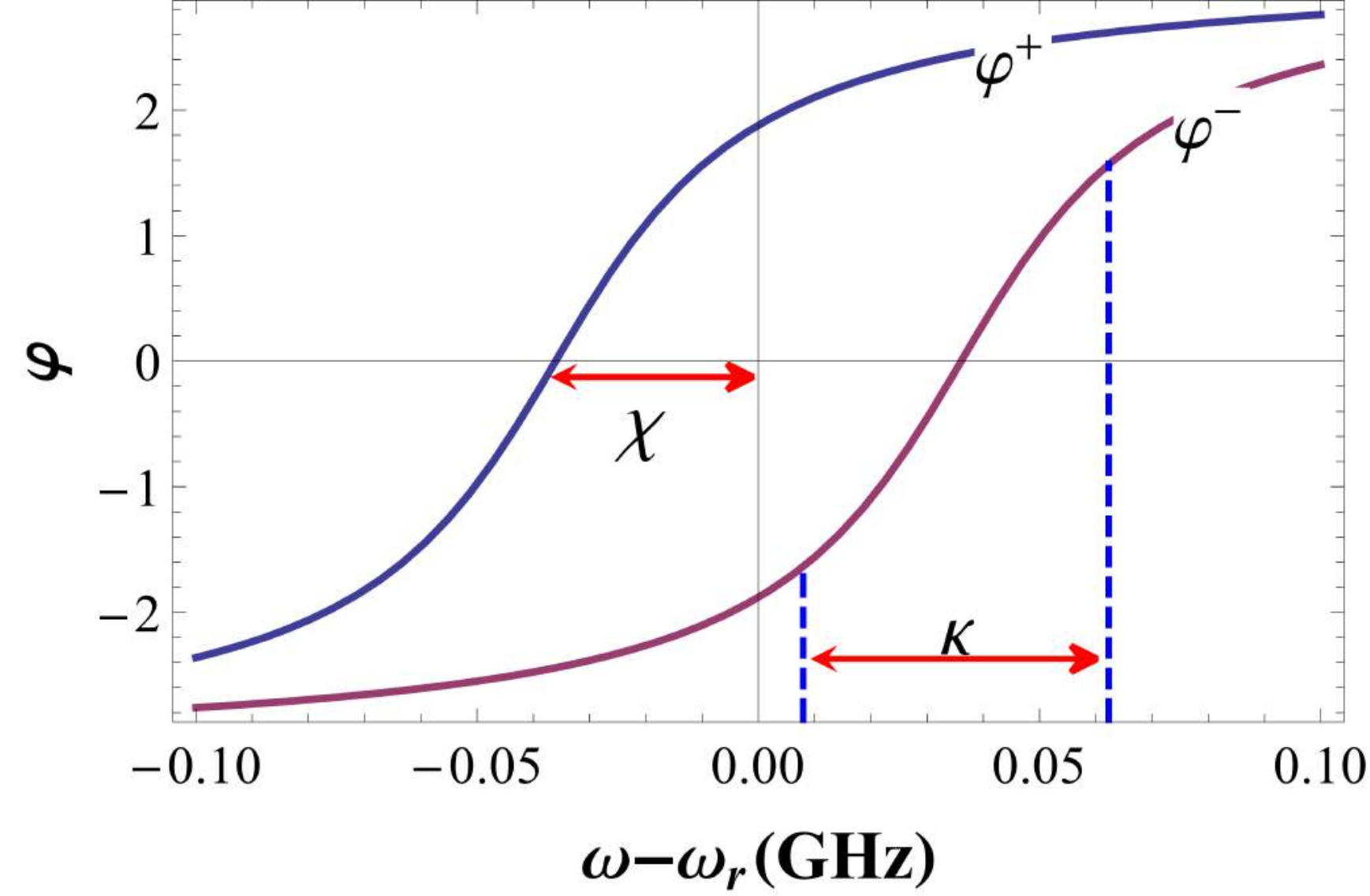}
\caption{Frequency dependence of phase shifts $\varphi^{\pm}(\omega-\omega_r)$ for $\kappa=52$MHz and $\chi=36$MHz. The shift between the two $\arctan$ functions equals $2\chi$.}
\label{fig:phaseshifts}
\end{figure}

Let us consider to what extent the frequency-dependence of the phase shift $\varphi(\omega-\omega_r)$ will play a role in the distinguishability of the output signals, see e.g. Fig.~\ref{fig:phaseshifts}. For $W \ll \kappa$ one can Taylor expand Eq.~(\ref{eq:phaseshift}) around $\omega_r$, i.e. 
\begin{eqnarray}
\varphi^{\pm}(\omega-\omega_r)=\pm 2 \arctan\left(\frac{2 \chi}{\kappa}\right) +\nonumber \\
(\omega-\omega_r)\frac{d \varphi^{\pm}}{d\omega}|_{\omega_r}+O\left(\frac{(\omega-\omega_r)^2}{\kappa^2}\right). 
\end{eqnarray}
Note that $\frac{d \varphi^{\pm}}{d \omega}|_{\omega_r}=\frac{4}{\kappa(1+(\frac{2\chi}{\kappa})^2)}$ is independent of whether the qubit is in the $\ket{0}$ or $\ket{1}$ state. This means that in the linear approximation where we neglect terms $O(\frac{(\omega-\omega_r)^2}{\kappa^2})$, the Gaussian envelope of the wave packet in time (or space) does not get distorted, but merely picks up a time delay $\sim 1/\kappa$ at the cavity that is the same for both $\pm$ signals. In this regime one expects the finite bandwidth to affect neither the signal nor the noise, see the expressions Eq.~(\ref{eq:sig_noise}) in \ref{sec:homo}. 

When we go beyond the first-order Taylor expansion, we can observe that the frequency-averaged phase shift, $\int d\omega |\alpha(\omega)|^2 |\varphi^{\pm}(\omega-\omega_r)|$, (which is relevant for $W \sim \kappa$) is {\em smaller} than $|\varphi^{\pm}(\omega-\omega_r)|$ due to the shape of the $\arctan()$ function, see Fig.~\ref{fig:phaseshifts}. This means that the phase-shift at $\omega=\omega_r$, which gives the optimal SNR, lies beyond the $\pi$ phase-shift point. We see this effect numerically in, for example, Fig.~\ref{fig:error_coherent} in Section \ref{sec:num_wig} for the standard coherent state measurement.

As argued before, we have to work under the restriction that the number of photons in the cavity at any given moment in time $\overline{n}(t)$ is well bounded below the critical number of photons $n_{crit}$. For a Gaussian microwave pulse
with bandwidth $W$ and a total of $n^{cav}_{\rm pulse}$ photons at the entrance to the cavity one has, using Eq.~(\ref{eq:ncav}) and (\ref{eq:gauss})
\begin{eqnarray}
\lefteqn{\overline{n}(t)=n_{\rm pulse}^{\rm cav}|{\rm FT}(f(\omega)h(\omega))|^2,} \nonumber \\ \; 
& & f(\omega)=\frac{\sqrt{\kappa}}{\kappa/2-i(\omega-\tilde{\omega}_r)},h(\omega)=\frac{e^{-(\omega-\omega_c)^2/W^2}}{(2\pi)^{1/4} \sqrt{W/2}}
\label{eq:ncavpulse}
\end{eqnarray}
where FT stands for the Fourier Transform and $n_{\rm pulse}^{\rm cav}$ is the total number of photons arriving at the cavity. This expression is also approximately valid for states arriving at the cavity in the interferometric schemes depicted in Fig.~\ref{fig:su11}: even though the coherent input state may get entangled with other modes or squeezed before (or after) arriving at the cavity, it remains almost a product state with respect to the frequency-dependent modes at all times. The linear transformation due to the amplifier mixes modes at frequency $\omega$ with $2 \Omega-\omega$ but at $\Omega=\omega_c$, $\alpha(2 \omega_c-\omega)=\alpha(\omega)$ with $\alpha(\omega)$ in Eq.~(\ref{eq:gauss}) and thus $\langle b_{in}^{\dagger}(\omega) b_{in}(\omega')\rangle \approx \beta^*(\omega)\beta(\omega')$ for some amplitudes $\beta(\omega)$. For such a state, $n^{cav}_{\rm pulse}$ in Eq.~(\ref{eq:ncavpulse}) will equal the gain $G$ of the amplifier before the cavity times the number of photons in the input probe.

\begin{table}
\caption{Representative Ranges of Relevant Parameters}
\centering
\begin{tabular}{ll}
Transmon qubit $\frac{\omega_q}{2\pi}$ and bare cavity frequency $\frac{\omega_r}{2\pi}$ & $3-11$GHz\\
Qubit $T_1/T_2$ time & $10-100\mu$sec \\
Cavity decay rate $\kappa/2\pi$ & $1-10$MHz \cite{sank:fast} \\
Pulse/Measurement time $T_{pulse}$ & $25-300$nsec \\
Dispersive shift $\chi/2\pi$ & $1-10$MHz\\
Jaynes-Cummings coupling $g/2\pi$ & $1-150$MHz \cite{sank:fast}\\
(phase-sensitive, degenerate) JPA amplifier gain $G$   & 30dB \cite{thesis:cast}\\
(phase-insensitive) JPC amplifier gain $G$ & 23dB  
\end{tabular}
\label{table:parameters}
\end{table}

\begin{figure}[htb]
\centering
\includegraphics[width=2.5in]{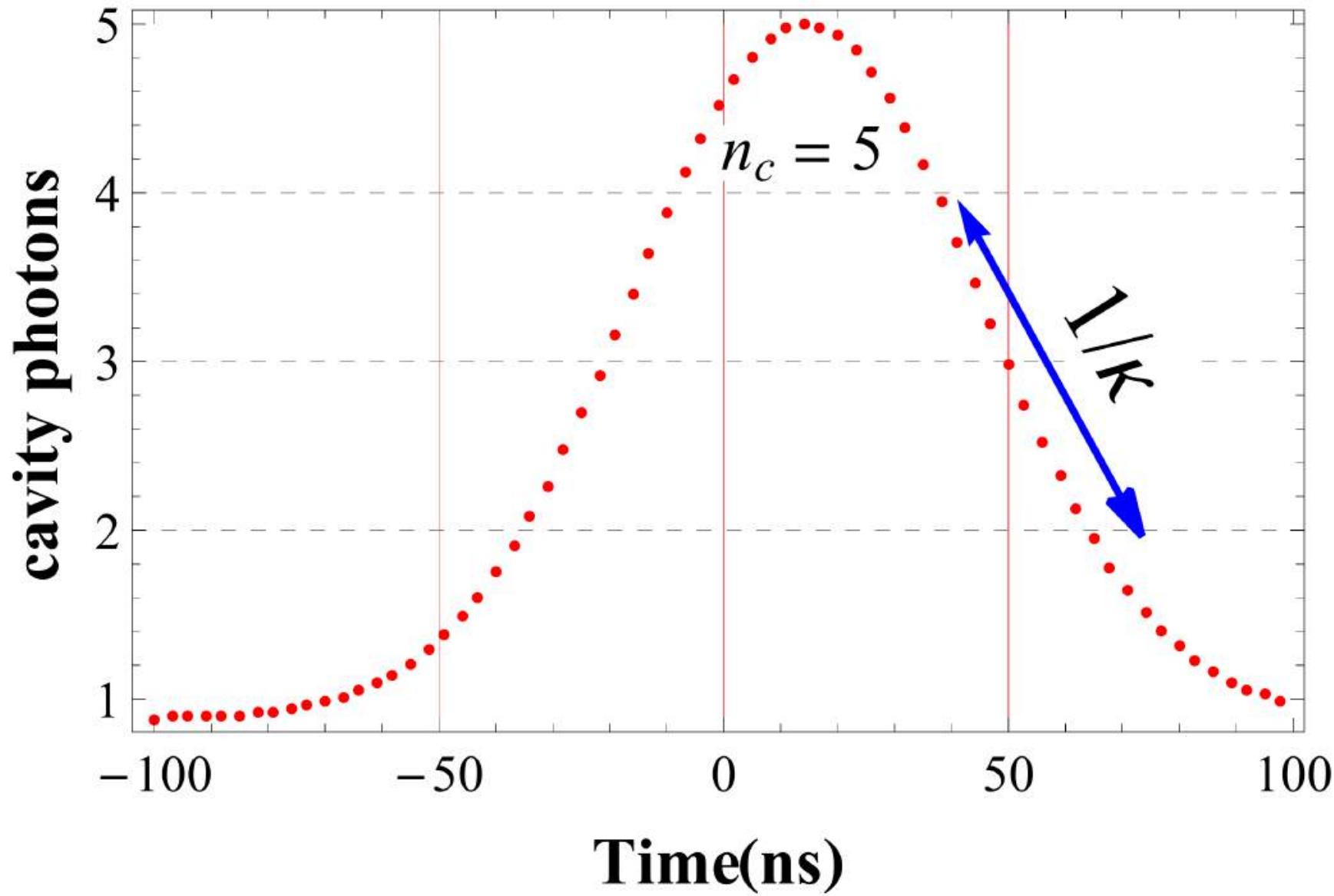}
\caption{Example of the number of photons in the cavity $\overline{n}(t)$ versus time $t$ for some representative choice of parameters $\Delta,g,\kappa, T_{pulse}, n_{pulse}$. The cavity initially is excited by a Gaussian pulse and the maximum number of photons inside the cavity $n_c$ stays well below the critical value set by the detuning $\Delta$ and the coupling $g$ at $n_{crit} \simeq 41.53$.}
\label{fig:res_photons}
\end{figure}

In Fig.~\ref{fig:res_photons} we use Eq.~(\ref{eq:ncavpulse}) to plot the instantaneous number of photons in the cavity $\overline{n}(t)$ for some illustrative parameters. We have taken a Gaussian pulse with $T_{pulse}=60$ns, cavity damping rate $ \kappa=1/25 $(ns), $\omega_{r}/2\pi=6.789$GHz and a qubit frequency at $\omega_q/2\pi=5.5$GHz \cite{sank:fast} so that the detuning equals $\Delta/2\pi=1.289 $GHz. By assuming a coupling strength $g/2\pi=100$MHz, the critical number of photons inside the cavity is $n_{crit}=\Delta^2/(4g^2) \simeq 41.53 $. The plot shows that the maximum number of photons $n_c=\max\bar{n}(t)$ of Eq. (\ref{eq:ncavpulse}) is at most 5 for a total number of input photons $n_{\rm pulse}=9$. For this choice of parameters one has $\frac{2 \chi}{\kappa}=2.43$ and a rather large value of $\frac{W}{\kappa}=0.83$.

A clear way of having a relatively short measurement time but remaining in the regime where the pulse is not distorted ($W/\kappa \ll 1$) is to have large cavity decay rate $\kappa$. However, via the coupling with the resonator mode the transmon qubit undergoes additional decoherence due to the Purcell effect \cite{book:haroche, koch+:purcell}. This can lead to a loss of the non-demolition character of the measurement as it speeds up the decay from $\ket{1}$ to $\ket{0}$. One can bound \cite{koch+:purcell} the Purcell-induced decoherence time $T_1 \leq \frac{\Delta^2}{\kappa g^2}=\frac{\Delta}{\kappa \chi}$ where $\Delta$ is the detuning. A larger $\kappa$ can thus be accommodated by increasing the detuning $\Delta$ (for identical $\chi$) leading to an increase in $n_{crit}$.  A route towards enhancing $\kappa$ without inducing additional decoherence was indicated in the experiment \cite{sank:fast}, which reports a fast single-qubit measurement with $T_{pulse} \sim 25$nsec.

\subsection{Amplification and Homodyne Measurement}
\label{sec:homo}

The microwave signal emerging from the cavity is amplified through, first, either a phase-sensitive or phase-insensitive amplifier at low ($\sim 30$mK) temperature, and subsequently, through some standard transistor amplifiers operating at higher temperatures, see Fig.~\ref{fig:expsu11}. The presence of these amplifiers does not impact which set-up or choice of parameters leads to an optimal SNR: the only requirement is that the signal coming into the sequence of amplifiers is already sufficiently strong so that the added noise of these amplifiers does not wash out any expected sensitivity enhancements.

We assume that the added noise of the Josephson parametric amplifiers is negligible and so their mode transformation correspond to the idealized ones of Eqs.~(\ref{eq:pa}) and (\ref{eq:dpa}) with a vacuum state at the idler port of the phase-insensitive parametric amplifier. A standard (e.g. transistor) linear amplifier can be modeled \cite{caves:noise} by mixing in continuum modes $c(\omega)$ (with $[c(\omega),c^{\dagger}(\omega')]=\delta(\omega-\omega')$) which are assumed to be in a thermal state at (an effective) temperature $T$ as follows:
\begin{eqnarray}
a_{amp}(\omega)=\sqrt{G_{\rm amp}}a_{1,out}(\omega)+\sqrt{G_{\rm amp}-1} \,c(\omega),
\end{eqnarray}
where $a_{1,out}(\omega)$ is the outgoing frequency mode in the various schemes in Fig.~\ref{fig:su11}. For a thermal state one has 
\begin{eqnarray}
\langle c^{\dagger}(\omega)c(\omega')\rangle=\overline{n}_T(\omega) \delta(\omega-\omega'), \\
\langle c(\omega) \rangle=0, \;\langle c(\omega) c(\omega') \rangle=0,\nonumber
\end{eqnarray}
where the number of photons at frequency $\omega$ is equal to $\overline{n}_T(\omega)=\frac{1}{e^{\hbar \omega/k_B T}-1} \approx \frac{k_B T}{\hbar \omega}$ for $k_B T \gg \hbar \omega$. Thus 
\begin{eqnarray}
\frac{(\Delta p_{amp}(\omega))^2}{G_{\rm amp}} \equiv (\Delta p_{1,out}(\omega))^2+A_N,\nonumber \\
= (\Delta p_{1,out}(\omega))^2+(1-G_{\rm amp}^{-1})\left(\overline{n}_T(\omega)+\frac{1}{2}\right), 
\label{eq:ampnoise}
\end{eqnarray}
where the defined $A_N$ is the added noise number. For the Josephson parametric amplifier one has $A_N < 1.7$ quanta while for the JPA in \cite{thesis:cast} $A_N$ is reported to be 0.23, below $\frac{1}{2}$ (this is a phase-sensitive amplifier). When the added noise number is considerably above $\frac{1}{2}$ we may thus associate an effective noise temperature $T_N \approx \frac{\hbar \omega \overline{n}(\omega)}{k_B} \approx  \frac{\hbar \omega A_N}{k_B}$ with it. The added noise is largely set by the first amplifier in the chain: for the HEMT amplifier (at operating temperature $T=5-10$K and frequency of $\frac{\omega}{2\pi}=4 -10$GHz one has $A^{\rm HEMT}_N \approx 20-30$ \cite{cast2008}. In order for the relative noise contribution from the HEMT to be small, the total number of photons in the outgoing signal (after the last PA or DPA) should thus be more than $20-30$. 

At the room-temperature output (see Fig.~\ref{fig:expsu11}), the chain of amplification thus produces an essentially classical stochastic voltage signal $V(t)\propto\frac{-i}{\sqrt{2}} (\alpha_{out}(t)-\alpha_{out}^*(t))$ with expectation $\langle V(t) \rangle$ and stochastic noise correlator $\Delta^2 V(t,t')\equiv\langle V(t) V(t') \rangle-\langle V(t) \rangle \langle V(t') \rangle$.  The last step in the quantum measurement chain is the measurement of this time-dependent voltage, which is usually referred to as a `homodyne measurement'. We assume that a single measurement outcome $\pm$ will be deduced after time $T_m$. This measurement time $T_m \sim T_{pulse}+O(\frac{1}{\kappa})$ when $W \sim \kappa$ so that we also catch the late-incoming photons. 

The homodyne measurement in practice means the mixing of the signal with a reference signal and the application of a low-band pass filter in order to eliminate the fast-oscillating behavior of $V(t)$ (or different quadratures). This homodyne measurement thus differs from the standard quantum optics technique in which a homodyne measurement of a weak quantum signal is realized by mixing it (on a partial beam splitter) with a high-amplitude local oscillator~\cite{foot3}. The output of the microwave homodyne measurement for the $p$-quadrature is the time-averaged signal
\begin{equation}\label{eq:sign}
\langle p_{out}\{T_m\}\rangle \equiv \sqrt{4 \pi} \int_{-\frac{T_m}{2}}^{\frac{T_m}{2}}dt\,\mathrm{cos}(\omega_c t)\, \langle V(t) \rangle, 
\end{equation} 
where $\omega_c$ is chosen to be equal to the carrier frequency of the incoming pulse equal to $\omega_r$. In a typical experiment \cite{hatridge+:back} a stochastic signal $p_{out}\{\delta t\}$ is obtained for shorter time intervals $\delta t \ll T_m$, but we assume here that one takes the sum over this entire data record $p_{out}\{\delta t\} \delta t$ and obtains one random variable $p_{out}\{T_m\}$ with mean as in Eq.~(\ref{eq:sign}). The noise on this signal is given by 
\begin{eqnarray}
\lefteqn{(\Delta  p_{out}\{T_m\})^2\equiv } \nonumber \\
& & 4 \pi \int_{-\frac{T_m}{2}}^{\frac{T_m}{2}}dt\int_{-\frac{T_m}{2}}^{\frac{T_m}{2}}dt'\mathrm{cos}(\omega_c t)\,\mathrm{cos}(\omega_c t') \Delta^2 V(t,t'). 
\label{eq:noise_exp}
\end{eqnarray}

In order to evaluate these expressions for the various set-ups described in Fig.~\ref{fig:su11}, let the modes $a_{1,out}(\omega)$ (with quadrature $p_{out}(\omega) \equiv p_{1,out}(\omega)$) describe the frequency-dependent output modes obtained from transforming $b_{out}(\omega)$ in Eq.~(\ref{eq:phase_out}) through the degenerate or non-degenerate parametric amplifiers, Eqs.~(\ref{eq:pa}) and (\ref{eq:dpa}) (overall we omit any time-delays that are picked up to due finite-speed propagation along the transmission lines). If we add the additional amplification and noise by the HEMT with gain $G_{\rm H}$, we can evaluate Eqs.~(\ref{eq:sign})-(\ref{eq:noise_exp}) to obtain the approximate expressions 
\begin{eqnarray}
\lefteqn{\langle p_{out}^{\pm}\{T_m\}\rangle \approx T_mG_{\rm H}^{\frac{1}{2}}\int_{-\infty}^{\infty} d\omega\,\mathrm{sinc}\left(\frac{T_m (\omega-\omega_c)}{2}\right)\langle p_{out}^{\pm}(\omega)\rangle}, \nonumber \\
& & (\Delta p_{out}^{\pm}\{T_m\})^2 \approx T_m^2 G_{\rm H}\int_{-\infty}^{\infty} d\omega \,\mathrm{sinc}^2\left(\frac{T_m(\omega-\omega_c)}{2}\right) \nonumber \\
 & & \left[(\Delta p^{\pm}_{out}(\omega))^2+(1-G_{\rm H}^{-1})\left(\overline{n}_T(\omega)+\frac{1}{2})\right)\right]. 
\label{eq:sig_noise}
\end{eqnarray}
When we evaluate these expressions, we will use values typical from the literature, viz., $\overline{n}_T=25$ and 
$G_{\rm H}=30.1$dB, so that $A^{\rm HEMT}_N=24.7$. To arrive at  Eq.~(\ref{eq:sig_noise}) we have neglected the terms proportional to $e^{\pm i(\omega+\omega_c)t}$ assuming that these fast-rotating terms average out because of the time integration. Note that the signal strength will increase for small $T_m$ but then saturate once all photons in the pulse have been processed. In calculating the noise we also use the fact that both the outgoing signal, as well as the thermal state that is mixed in with the signal by the HEMT amplifier, are product states with respect to the frequency modes. Note that the probability distribution of this quadrature random variable is a Gaussian distribution as all states in the protocol --- coherent, squeezed and thermal --- are Gaussian states.
It is interesting to see what happens in Eq.~(\ref{eq:sig_noise}) when the measurement time $T_m$ becomes too large while we keep $T_{\rm pulse}$ fixed. The signal becomes constant as all photons have been processed, but the noise continues to grow $(\Delta p_{out}^{\pm})^2 \sim T_m$ so that the SNR goes to zero. 

\subsection{Signal-to-Noise Ratio and Measurement Error Probability}
\label{sec:error}

The quality of the quantum measurement, given a fixed measurement time $T_m$, can be expressed indirectly using a signal-to-noise ratio SNR  (Eq.~(\ref{eq:SNR})), and more directly through a measurement error probability. If the qubit is in the $\ket{0}$ (resp. $\ket{1}$) state, the outgoing signal distribution $P_{\pm}(x=p_{out}\{T_m\})$ is a Gaussian distribution $P_{\pm}(x)=\frac{1}{\sigma_{\pm} \sqrt{2 \pi}} \exp(\frac{-(x-\mu_{\pm})^2}{2\sigma_{\pm}^2})$ with mean $\mu_{\pm}=\langle p_{out}^{\pm}\{T_m\}\rangle$ and standard deviation $\sigma_{\pm}=\Delta p_{out}^{\pm}\{T_m\}$ such that $\mu_+ \geq 0$. As argued in Section \ref{sec:schemes}, the standard deviations $\Delta p_{out}^{\pm}\{T_m\}$ are not necessarily identical for arbitrary choice of phases in the SU(1,1) interferometer, but we will only choose parameters such that $\Delta p_{out}^+ =\Delta p_{out}^-$. Assuming that the qubit has an arbitrary long lifetime compared to the measurement time $T_m$, one chooses a mid-way threshold value $\nu=\mu_-+\frac{|\mu_+-\mu_-|}{2}$ (for $
\mu_- < \mu_+$) such that when 
$x < \nu$ we decide for outcome `$-$' or $\ket{1}$, while for $x > \nu$ we decide `$+$' or $\ket{0}$ (for a qubit with a finite lifetime one should bias this threshold value, see \cite{gambetta+:protocols}). The probability for an incorrect measurement conclusion is equal to $P_{\rm error}={\rm Prob}(\mbox{infer}-|+) {\rm Prob}(+)+{\rm Prob}(\mbox{infer}+|-){\rm Prob}(-)$ and we will assume an equal probability for $\pm$, ${\rm Prob}(\pm)=\frac{1}{2}$. Using ${\rm Prob}(-|+)={\rm Prob}(+|-)=P_{-}(x \geq \nu)=\frac{1}{2}{\rm erfc}\left(\frac{|\mu_+-\mu_-|}{2\sqrt{2}\sigma}\right)$ gives
\begin{eqnarray}
P_{\rm error}(T_m)=\frac{1}{2}\mathrm{erfc}\Big[\frac{|\langle p_{out}^+\{T_m\}\rangle-\langle p_{out}^-\{T_m\} \rangle|}{2 \sqrt{2} \Delta p_{out}^{\pm}\{T_m\}}\Big] & & \nonumber \\  =\frac{1}{2}\mathrm{erfc}\left(\frac{{\rm SNR}}{\sqrt{2}}\right),& &
\label{eq:error}
\end{eqnarray}
with the expressions in Eq.~(\ref{eq:sig_noise}). This error probability does not say to what extent the measurement also projects the qubit onto the $\ket{0}$ or $\ket{1}$ state given the measurement outcomes; this additional information could be obtained through a stochastic master equation analysis as in \cite{gambetta+:zeno}. Note also that the statistical reasoning leading up to these expressions is a shorthand for the actual situation, as the qubit is not generally in either the $\ket{0}$ or $\ket{1}$, but can be in an arbitrary superposition $a \ket{0}+b \ket{1}$,  Again, a stochastic master equation analysis would give a more complete description of the gradual `collapse' of the qubit wavefunction.\\

It is common to include a `fudge' measurement inefficiency factor $\eta < 1$  in the final measurement error probability to account for the fact that not all photons in the measurement pulse contribute to the outgoing signal (as they get reflected etc.). For example, for a cavity with two ports each with decay rate $\kappa_{in}$ and $\kappa_{out}$ of which only the out port is monitored, one has \cite{gambetta+:zeno} $\eta=\frac{\kappa_{out} \eta_{det}}{\kappa_{in}+\kappa_{out}}$ where $\eta_{det}$ is some overall efficiency of detecting the photons at the output. In \cite{delange+:eta_determin} the authors determine an overall measurement efficiency of $\eta \approx \frac{1}{2}$. As the loss of photons could be abstractly modelled as the presence of an additional beam-splitter somewhere in the measurement chain, a good approximation to the modified error probability is then $P_{\rm error}=\frac{1}{2}{\rm erfc}\left(\sqrt{\frac{\eta}{2}}{\rm SNR}\right)$ which we will use in our numerical evaluations. However one expects that a more detailed modeling of photon loss inside the interferometer versus photon loss at the outgoing/ingoing ports, would modify the SNR in different ways. Loss inside the SU(2,2) interferometer would lead to both a loss in signal (similar as for the coherent measurement) as well as a loss in the entanglement of the two-mode squeezed state which comes out of the first PA, thus reducing the advantage of the interferometer. We leave the calculation of the various SNRs of such lossy interferometers as future work.

\section{Numerical Evaluation of Schemes}
\label{sec:num_wig}

Overall, it seems daunting, if not impossible, to be able to experimentally determine the precise values of all the physical parameters which play a role in the SNR. It assumes these values can be determined independently of each other, at least these should only depend on parameters whose values we already know with high accuracy. But for a theoretical study we are not faced with this conundrum and we can consider the sensitivity enhancement that one may be able to achieve given realistic values of these parameters. This is what we do in this section. 

\begin{figure}[htb]
\centering
\includegraphics[width=3in]{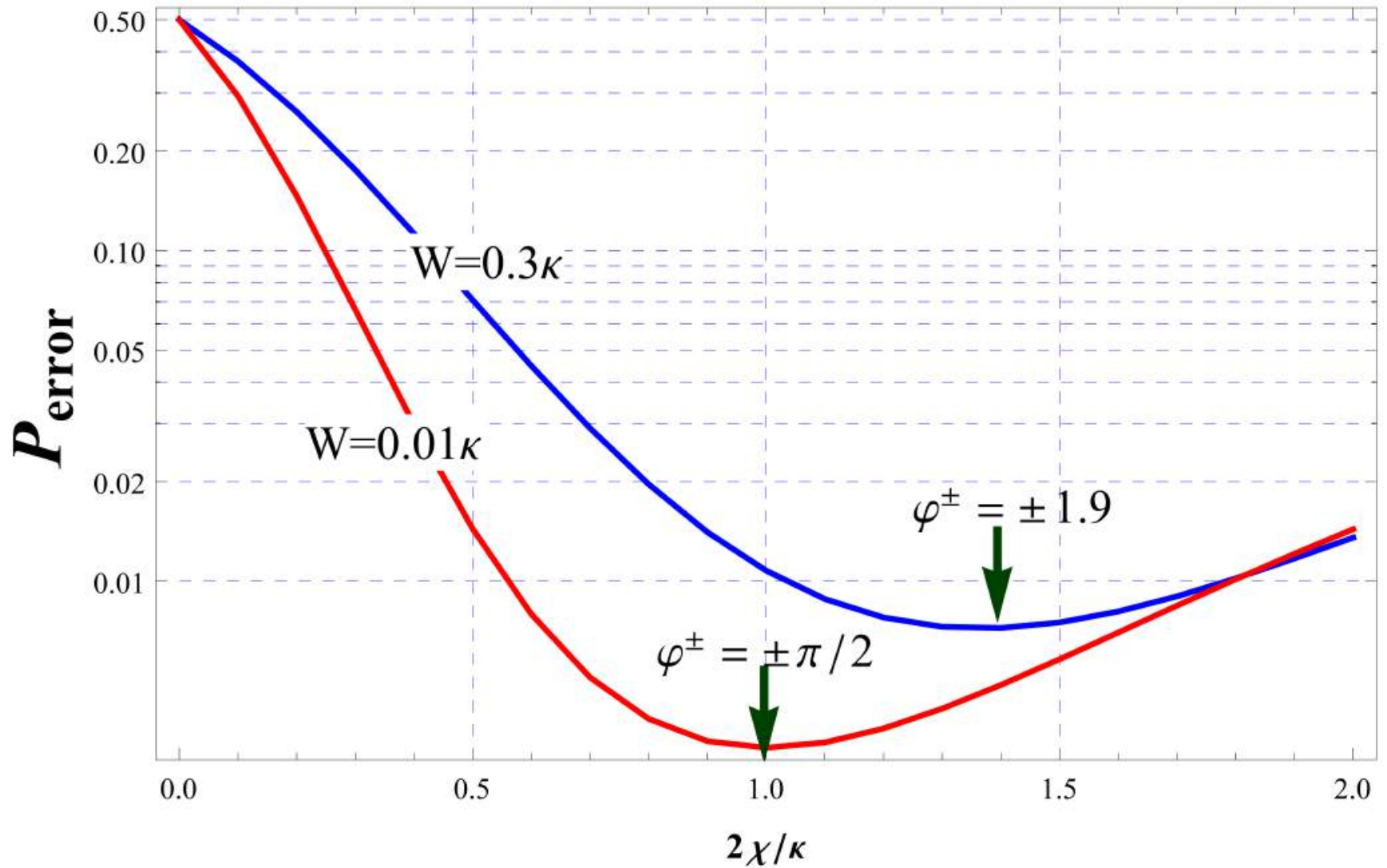}
\includegraphics[width=3in]{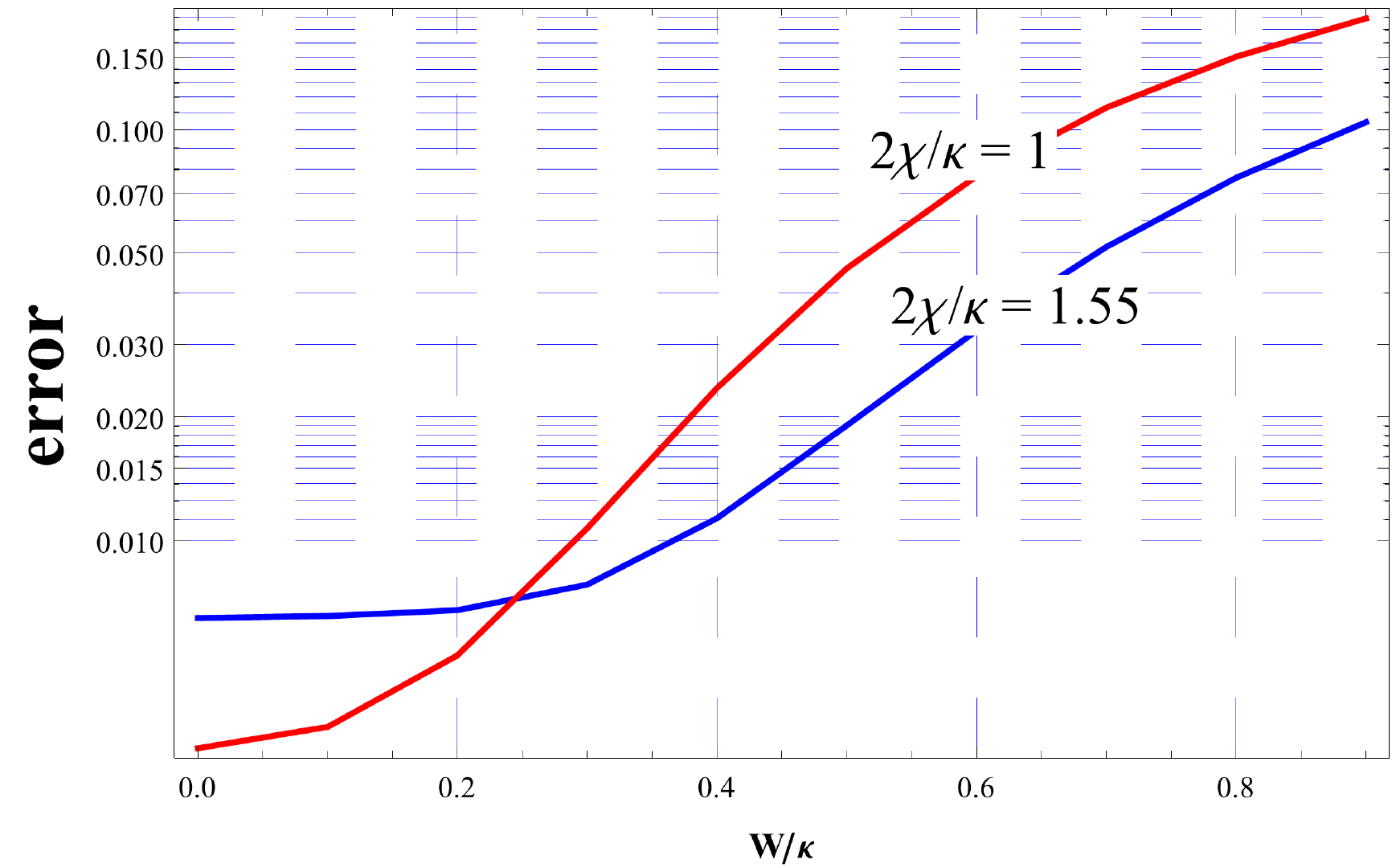}
\caption{(a) The probability of error $P_{\rm error}$ of the standard coherent measurement scheme versus $\frac{2\chi}{\kappa}$.  Other parameters in this plot (which do not directly affect where the error probability is minimal) are $\omega_r/2\pi=6.789 $GHz, $\omega_q/2\pi=5.5$GHz, and $T_m=1.2T_{pulse}$, $T_{\rm pulse}=2/W$, $n_{pulse} =9$. (b) $P_{\rm error}$ versus normalised pulse width $W/\kappa$ for different values of phase shift (determined by $2\chi/\kappa$). }
\label{fig:error_coherent}
\end{figure}

We first observe that in the standard coherent scheme, the finite pulse $T_{\rm pulse}$ affects the values of $\varphi^{\pm}$ for which the SNR is maximized, see Fig.~\ref{fig:error_coherent}. For a pulse width $W=0.01\kappa$, the minimal error probability is found at the $\pi$-phaseshift point of $2\chi/\kappa\approx 1$, but for $W=0.3\kappa$ the minimal error probability lies at $2\chi/\kappa \approx 1.4$. We do not limit the number of photons in the cavity, but it is noticeable that even for $W=0.3\kappa$, the minimal error probability is achieved for phase shifts $|\varphi^+(0)-\varphi^-(0)| > \pi$, see Eq.~(\ref{eq:phaseshift}). This demonstrates that we should include the multi-mode nature of the pulse in our analysis.

\subsection{Comparison between SU(1,1) Interferometer and coherent readout: single mode results}

\begin{figure}[ht]
\centering
\includegraphics[width=2.5in]{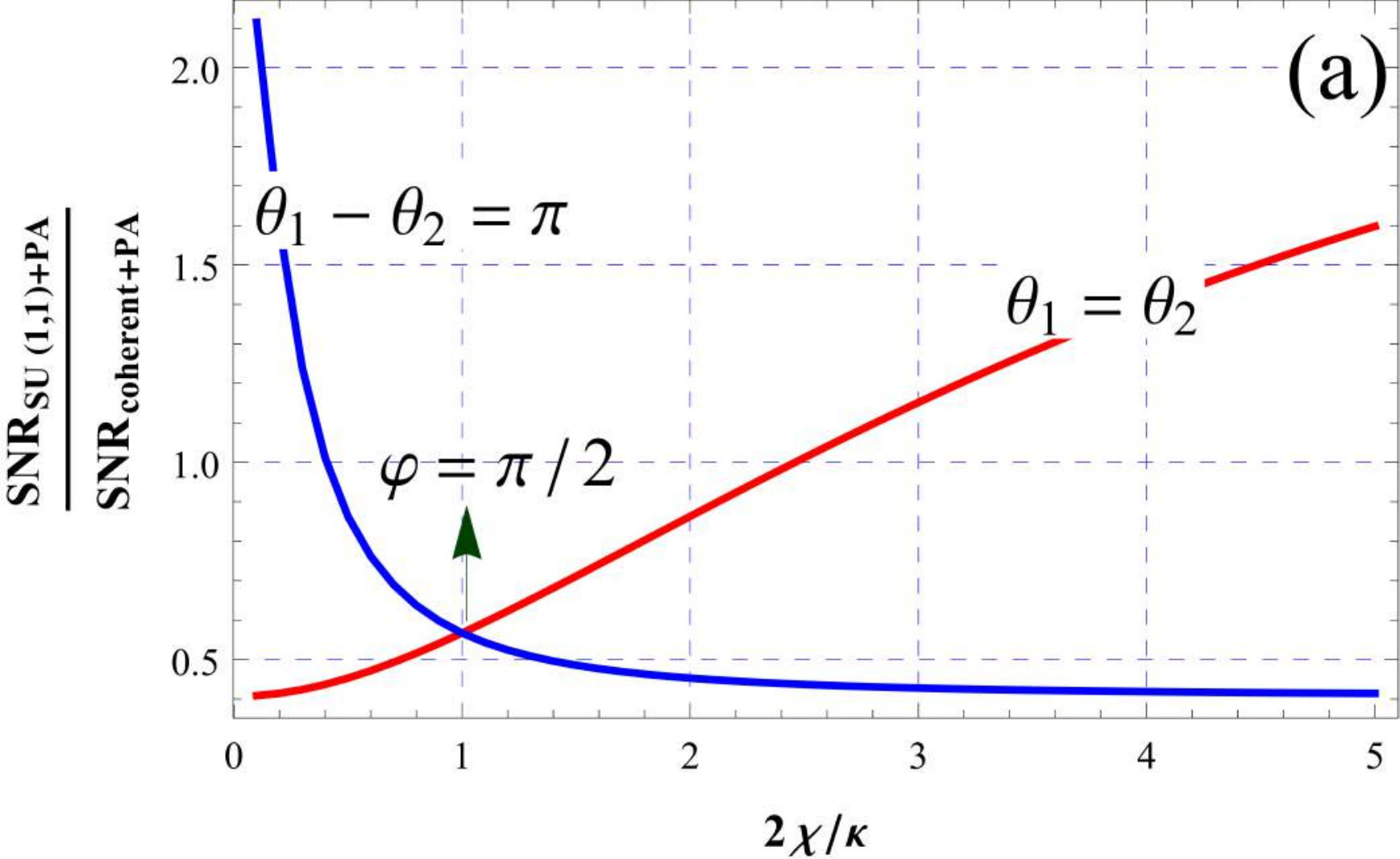}
\includegraphics[width=2.5in]{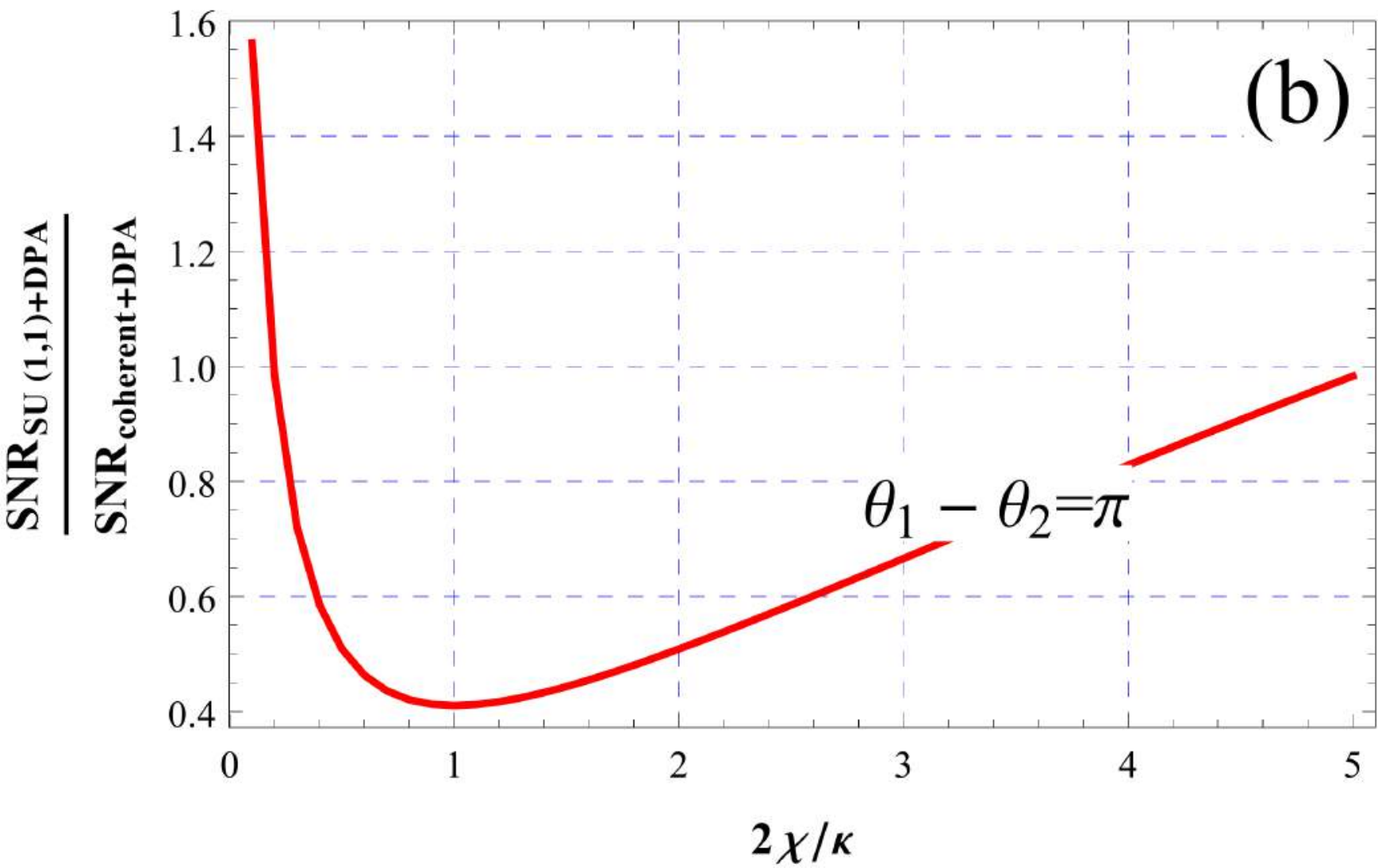}
\includegraphics[width=2.5in]{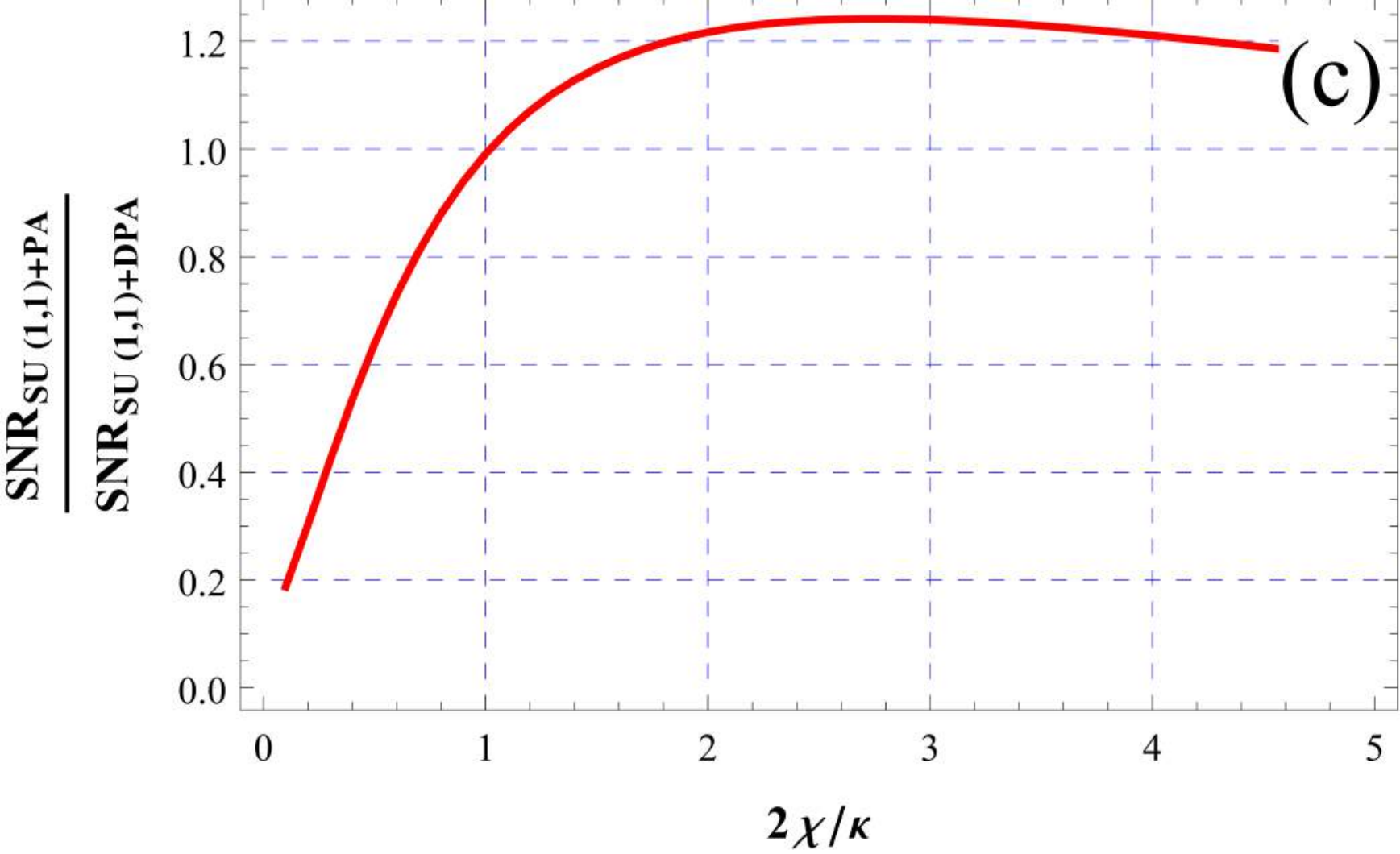}
\caption{(a) Comparison between $\rm SNR_{SU(1,1)+PA}$ and $\rm SNR_{coherent+PA}$ for two different relative phases $ \theta_2-\theta_1=\pi $ and  
$ \theta_2=\theta_1 $. (b) Comparison between $\rm SNR_{SU(1,1)+DPA}$ and $\rm SNR_{coherent+DPA}$ for the optimal phase choice $ \theta_2-\theta_1=\pi$. (c)Comparison between $\rm SNR_{SU(1,1)+DPA}$ (with the optimal phase choice $ \theta_2-\theta_1=\pi$ ) and $\rm SNR_{SU(1,1)+PA}$ (with the optimal phase choice $ \theta_2=\theta_1)$. In all scenarios we assume  the optimal $G_1=3.12$dB, $G_2=20$dB (a practical maximum amplifier gain),  and the number of photons in the cavity is at most 5. In the case of PA-based SU(1,1) interferometer the photon flux (see Eq. (\ref{eq:ncav}) and Appendix A) before the first PA is $31.23 \kappa$ and the photon flux after the first PA but before the cavity is $F_t=32.5 \kappa$ which is equal to the photon flux before the cavity for the coherent schemes. For the DPA-based interferometer the photon flux before the cavity is $21.7 \kappa$ which is amplified to $32.5 \kappa$ after the first DPA.}
\label{compareSNR}
\end{figure}

We first consider the idealized situation, as discussed in Section \ref{sec:schemes} where $T_{\rm pulse}$ is large so that one can make a single-mode approximation. We can take the measurement time $T_m$ to be sufficiently long so that the system reaches steady-state and a constant flux of photons is arriving at the output. Under these assumptions, the probability of error of the coherent state readout~(coherent+PA)  and the two-mode SU(1,1) interferometer~(SU(1,1)+PA) are given by  Eq.~(\ref{eq:SNRcoh}) and Eq.~(\ref{eq:SNRsu11_1}), respectively. In Fig.~\ref{compareSNR} we plot the ratio of these two SNRs against $2\chi/\kappa $ and for two different values of the PA phase differences $ \theta_1-\theta_2 $. Here we take the number of photons in the cavity $\overline{n}(t) \leq 5$ and we use $G_1=3.12$dB.  Here, and below, we take $G_2=20 $dB; larger amplification by the second stage is always improves the SNR, so we take the largest value that is easily attainable in currently used PAs. As expected, for $\theta_1-\theta_2 =\pi$ the two-mode SU(1,1) interferometer shows higher measurement accuracy as compared to the coherent state readout with PA for very small values of $ 2\chi/\kappa \ll 1$. However, this is not the regime that we are interested in since the signal, being proportional to $ \mathrm{sin(\varphi)}$, will be very small for these values of $ 2\chi/\kappa$. On the other hand, when $ \theta_1=\theta_2 $ the two-mode SU(1,1) interferometer shows a better result as compared to the coherent state readout for large phase shift $ 2\chi/\kappa $. We will focus our analysis on rather large values of $2\chi/\kappa $ where we have a significant signal.

Similarly, we can compare the SNR of the single-mode SU(1,1) interferometer ($\rm SNR_{ SU(1,1)+DPA}$ in Eq.~(\ref{eq:SNRsu11_2})) with the SNR of a coherent pulse which is amplified using a DPA ($\rm SNR_{ coherent+DPA}$), see Eq.~(\ref{eq:cohDPA}). As one can see, the DPA-based SU(1,1) interferometer gives a higher SNR as compared to the coherent DPA-based readout when the phase shifts are very small $ 2\chi/\kappa \ll 1$ or very large $ 2\chi/\kappa>2.9$. Thus Fig.~\ref{compareSNR} shows that the optimal scenario for qubit state readout is the two-mode SU(1,1) interferometer for the relevant values of $2\chi/\kappa$. In the next section we present the multi-mode features of this scenario and we will omit the multi-mode numerics of the other scenarios.
 
\subsection{Two-Mode SU(1,1) Interferometer: Multi-mode Numerics}

Now we consider the more realistic scenario in which a coherent pulse with a total of $n_{\rm pulse}$ photons and bandwidth $W$ is provided as input to the SU(1,1) interferometer, as shown in Fig.~\ref{fig:su11}(a). This pulse first amplifies(deamplifies) at the first PA (with gain $G_1$ and phase $\theta_1$), then interacts with the qubit in the cavity, picks up a frequency-dependent phase factor as in Eq.~(\ref{eq:phaseshift}), and is subsequently amplified by the second PA and then by the amplifiers at higher temperatures (see the setup in Fig.~\ref{fig:expsu11}).

\begin{figure}[ht]
\centering
\includegraphics[width=3in]{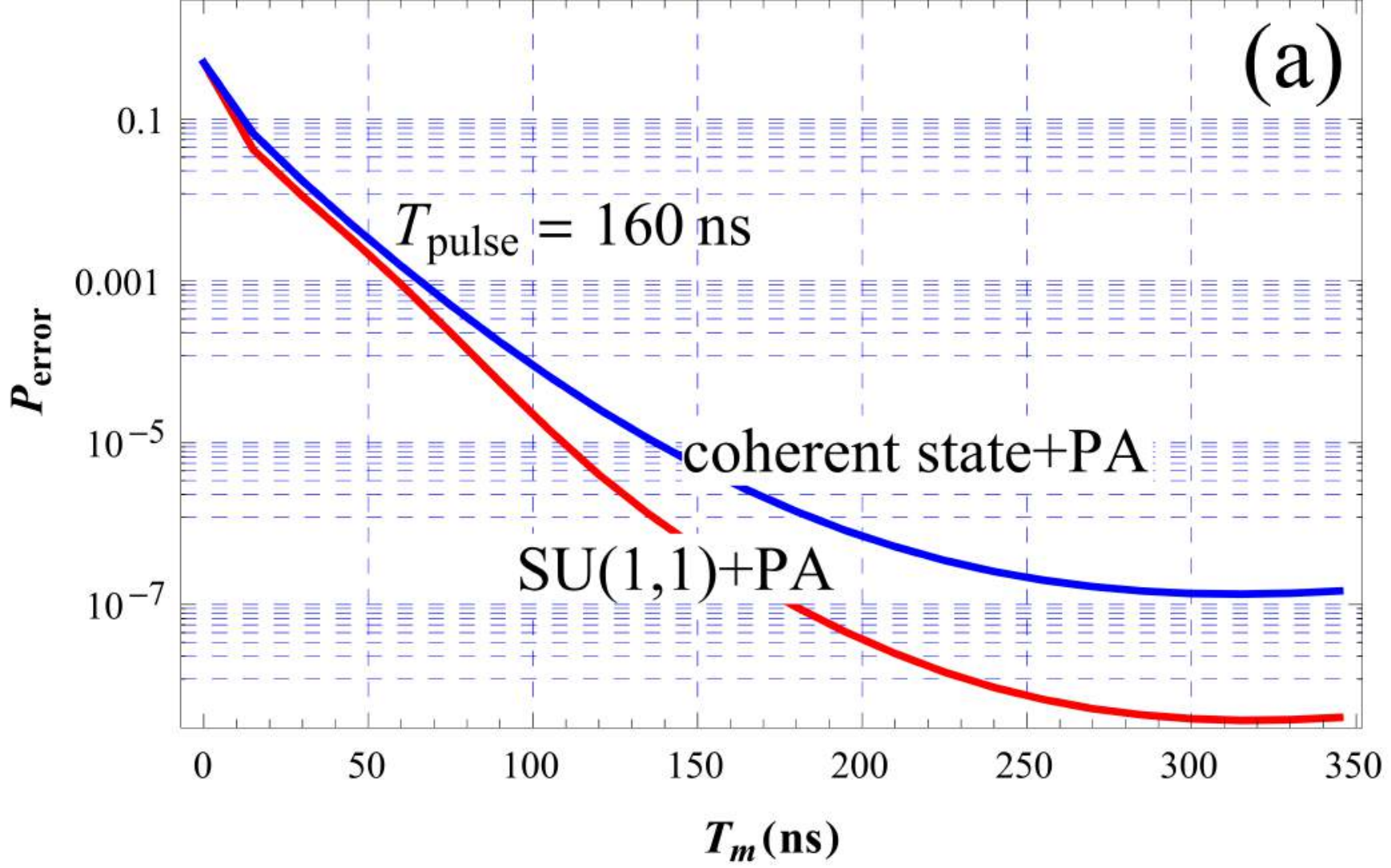}
\includegraphics[width=3in]{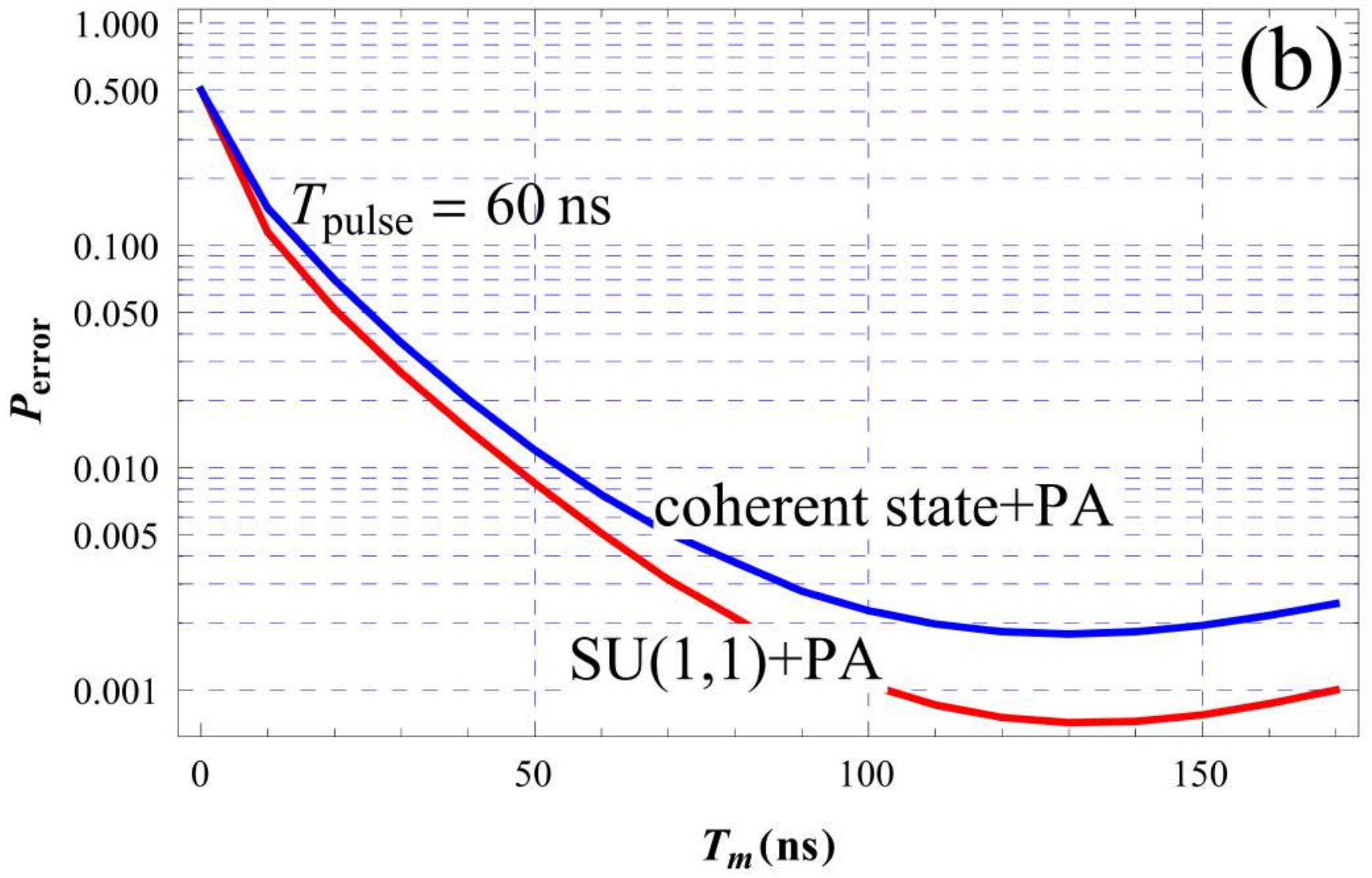}
\caption{Central result of the paper: The probability of error versus measurement time $T_m $ (a) for a pulse with time duration $ T_{pulse}=160ns $ where the optimum total number of photons in the pulse before first PA is $n_{\rm pulse}=58.98 $. (b) For a pulse with time duration $ T_{pulse}=60ns $ where the optimum total number of photons in the pulse before the first PA is $n_{\rm pulse}=19.36$.}
\label{snr_noise11}
\end{figure}

In Fig.~\ref{snr_noise11} we plot the probability of error $P_{\rm error}$ versus the measurement time $T_m$ for the two-mode SU(1,1) interferometer and for a coherent state readout, using the expressions for the signal and the noise in Eq.~(\ref{eq:sig_noise}), Eq.~(\ref{eq:SNR}) and the expression for the probability of error $P_{\rm error}=\frac{1}{2}{\rm erfc}\left(\sqrt{\frac{\eta}{2}}{\rm SNR}\right)$. Thus at $T_m=0$ with no photons at the output one has $P_{\rm error}=\frac{1}{2}$ while for large $T_m/T_{\rm pulse} \rightarrow \infty$, $P_{\rm error} \approx \frac{1}{2}{\rm erfc}( c T_m^{-1/2}) \rightarrow \frac{1}{2}$ for some constant $c$. We assume the experimentally realizable parameters \cite{sank:fast} $1/\kappa=25 $(ns), $ \chi/2\pi=7.7 $MHz, $2\chi/\kappa=2.43 $, $\eta=0.5 $,  $\omega_r/2\pi=6.789 $GHz, $\omega_q/2\pi=5.5$GHz. We have obtained these data by first fixing the maximum number of photons in the cavity to be 5, using Eq.~(\ref{eq:ncavpulse}). Given a value for $n_{\rm pulse}$ and the other parameters, this fixes the gain of the first amplifier $G_1$. We then consider for what value of $n_{\rm pulse}$ (recall that this is the total number of input photons, see Eq. (\ref{defpul})) the SNR is maximized and present the optimal value. The figure of merit that is thus held constant in comparing a coherent read-out and the SU(1,1) interferometer is thus the maximum number of cavity photons and the corresponding number of photons that is arriving at the cavity.

However, we consider two different values for the pulse duration. In Fig. \ref{snr_noise11}(a) we consider $T_{\rm pulse}=160ns $ where the optimal total number of photons in the pulse before first PA is $n_{\rm pulse}=58.98 $~(corresponding to $G_1=0.431$dB). In Fig. \ref{snr_noise11}(b) we assume  $T_{\rm pulse}=60ns $ where the optimum total number of photons in the pulse before first PA is $n_{\rm pulse}=19.36 $~(with corresponding $G_1=0.222$dB). As the parameters are chosen such that $2\chi/\kappa>1 $, the phase difference between first and second PA is set to be equal ($\theta_2=\theta_1$, as predicted by Fig. \ref{compareSNR}(a)) in order to get the best results by using an interferometer. 

The greater relative advantage of the SU(1,1) scheme for the longer pulse (Fig.~\ref{snr_noise11}(a)) has a straightforward explanation using the single-mode analysis.  For the parameters of Fig.~\ref{snr_noise11}(a), the steady-state SNR for the SU(1,1)+PA scheme is about 1.15 times greater than the SNR for the coherent-state scheme (cf. Fig.~\ref{compareSNR}, but with different system parameters).  In steady state, both SNRs grow like the square root of time (cf. Eqs.~(\ref{eq:SNRcoh}) and (\ref{eq:SNRsu11_1})), so the ratios of the two error rates should go like ${\rm erfc}(1.15 c\sqrt{t}) /{\rm erfc}(c\sqrt{t})$ for some constant $c$.  This is a growing function of time, agreeing with the trend seen in going to longer pulse times (from Fig.~\ref{snr_noise11}(b) to Fig.~\ref{snr_noise11}(a)).

It should be observed that the dramatic advantage gained by using the parameters of Fig.~\ref{snr_noise11}(a) would not be attainable in current practice: $T_1$ relaxation of the qubits would need to be much longer in order for measurement error rates of $10^{-5}$ to be realistic.  The gain indicated for the shorter probe pulse of Fig.~\ref{snr_noise11}(b) should be attainable by present-day superconducting qubits, but the gain in error rate here is much more modest (a factor of 2).  We have sought for parameters for which an order of magnitude gain in error rate would attainable with practical present-day qubits, for example by assuming a critical cavity photon number larger than 5 (permitting larger $G_1$ and thus, presumably, more entanglement of the two beams of the SU(1,1) interferometer).  However, up to this point we have not found other parameters for which this desired gain would be achieved.  We plan further studies to explore more of the large parameter space of SU(1,1) operation. 

\begin{figure}[ht]
\centering
\includegraphics[width=2.5in]{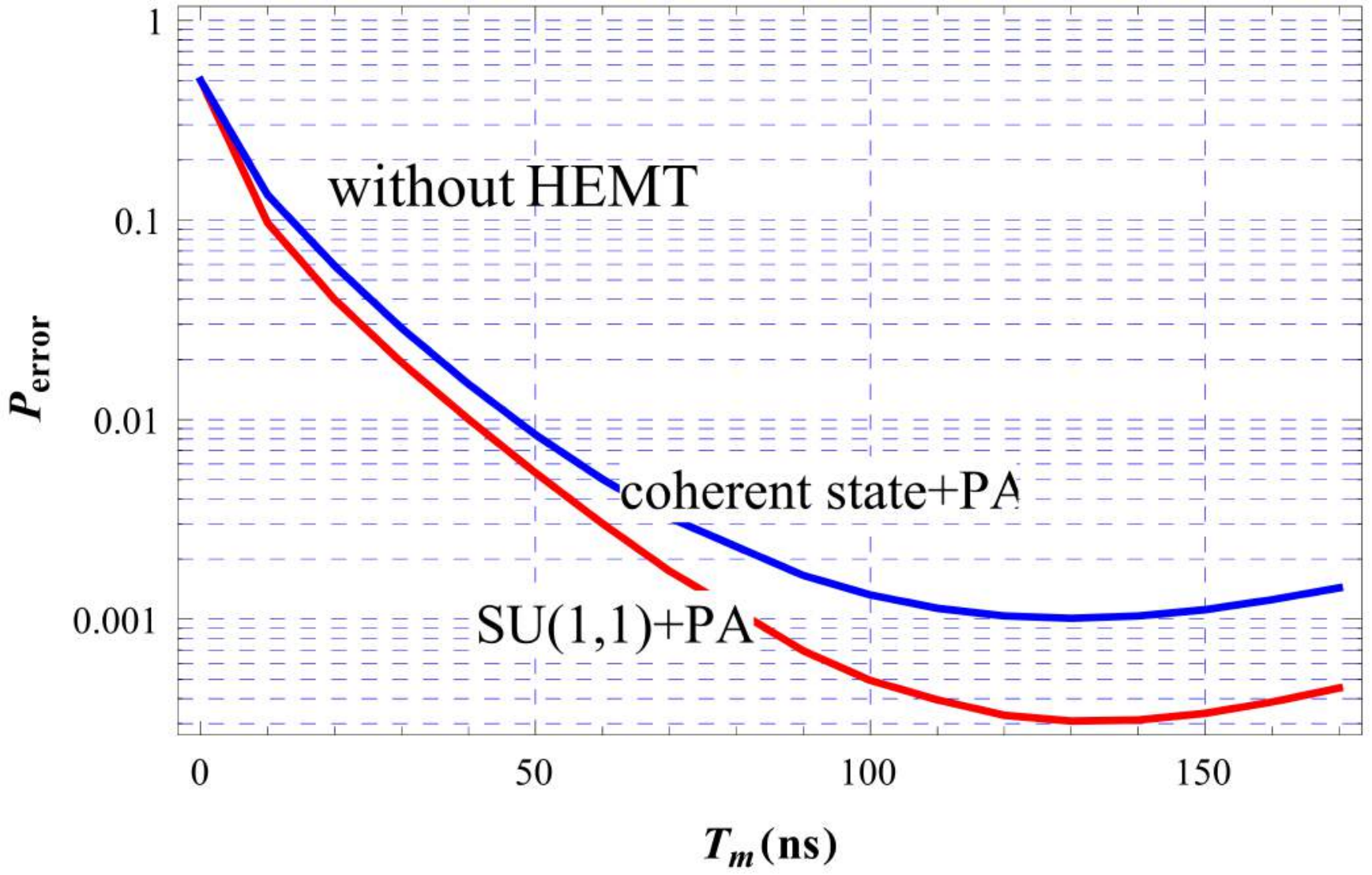}
\caption{ Same as Fig.~\ref{snr_noise11}(b), except that the assumed noisy post-amplification (with HEMT) is replaced by ideal amplification.   This quantifies the (small) amount by which the HEMT noise increases the error rate -- at the minimum, the change is from 0.0007 to 0.0003.  The relative performance of the coherent state vs. SU(1,1) schemes is essentially unchanged. 
}
\label{noHEMT}
\end{figure}

We finally show, in Fig.~\ref{noHEMT}, that there is a small but real degradation of the measurement due to the noisy HEMT post-amplification; the relative performance of the coherent-state vs. SU(1,1) schemes is unaffected by this degradation.

\section{Discussion}
\label{last}

In this paper we have considered the gains in measurement sensitivity for the measurement of a transmon qubit that is coupled to a microwave cavity using squeezers and parametric amplifiers, possibly in an interferometric setup.  None of these schemes for finite phase shifts $\varphi^{\pm}$ can claim to reach a Heisenberg limit, but we have demonstrated that an improvement in measurement fidelity for a given measurement time is possible. 

As an outlook for the future, we believe that it is worthwhile to consider the idea of feedback on the basis of partial homodyne measurement records both for the SU(1,1) interferometer as well as the single-mode SU(1,1) interferometer. The idea of feedback in this setting is different than in the usual setting of phase estimation in which one
biases the operating point of the SU(1,1) interferometer (the relative phases between the two parametric amplifiers) given the current precision with which the phase is known so as to be maximally sensitive to the remaining unknown bits \cite{YMK}. In a qubit measurement, the goal is to drive the qubit state as quickly as possible to either a
$\ket{0}$ or $\ket{1}$; if we gather an initial data record that suggests outcome $+$, one could bias the interferometer so that the SNR for the + signal becomes larger, but at the same time the SNR for the $-$ signal becomes smaller. Circuit QED techniques are certainly available for making the necessary fast, controllable changes of
propagation phase \cite{tune08}.  Whether this has the desired effect of realizing a faster projective measurement could be analyzed using stochastic master equations.

We can note that such feedback schemes do not give gains for a non-interferometric set-up such as a measurement with a squeezed or coherent probe. Even though the expression for the SNR in these non-interferometric setups does depend on the value of $\varphi^{\pm}$, shifting the carrier probe frequency $\omega_c$ to be close to a point $\varphi \approx 0$ will also bring one closer to resonance and hence lead to more photons in the cavity. In contrast, changing the relative phase of amplifiers before and after the probe has interacted with the cavity has no such effect. Another possible regime of interest is to consider probing the cavity far off-resonance with a sequence of very short pulses, possibly in an adaptive manner. Each pulse will pick up a small phase shift $\varphi^{\pm}$ at the cavity so that one can maximally benefit from using an interferometer. 

\section{Acknowledgements}
We would like to thank Eva Kreysing for working with us on preliminary results concerning the use of interferometers for measuring transmon qubits. We acknowledge funding through the EU via the SCALEQIT program.  DDV and SB are grateful for support through the Alexander von Humboldt Foundation.

\appendix

\section{Background on Modeling The Quantum Measurement Chain}
\label{sec:qmc_details}

Here we collect some mathematical relations and definitions pertaining to the description of a one-dimensional transmission line coupled to a (microwave) cavity \cite{clerk+:rmp} and review the input-output formalism \cite{book:walls_milburn, book:gardiner_zoller, blow+:cont}.

A semi-infinite transmission line can be described by a Hamiltonian $H_{\rm trans}=\int_{-\infty}^{\infty} dk\;\hbar \omega_k (b_k^{\dagger}b_k+\frac{1}{2})$ with $[b_k, b^{\dagger}_{k'}]=\delta(k-k')$ where $\omega_k=|k|v$ and the group (or phase) velocity is $v=\frac{1}{\sqrt{lc}}$ (dispersionless medium). Here $l$ ($c$) are the inductance (capacitance) per unit length of the line ($Z_c=\sqrt{\frac{l}{c}}$ is the characteristic impedance of the line). Note that the operators $b_k$ have units of ${\rm m}^{1/2}$ in this continuum limit. One can obtain this description through solving the one-dimensional wave equation $\frac{\partial^2 \Phi(x,t)}{\partial t^2}-v^2\frac{\partial^2 \Phi(x,t)}{\partial x^2}=0$ for the flux variable $\Phi(x,t)$ along the line and expanding it in normal modes labeled by a wave number $k$. The flux variable $\Phi(x,t)$ determines the local voltage $V(x,t)$ and local current density $I(x,t)$ through $\frac{\partial \Phi(x,t)}{\partial t}=V(x,t)$ and $I(x,t)=-\frac{1}{l} \frac{\partial \Phi(x,t)}{\partial x}$. The local voltage operator $V(x,t)$ at a point $x$ on the line is a Heisenberg operator and equals 
\begin{equation}
V(x,t)=-i \sqrt{\frac{1}{4 \pi c}} \int_{-\infty}^{\infty} dk \sqrt{\hbar \omega_k} \left(b_k(t)e^{i kx}-b_k^{\dagger}(t)e^{-i k x}\right), \nonumber  
\end{equation}
with $b_k(t)=b_k e^{-i \omega_k t}$. If we split the integral over wave numbers $k$ into a right-travelling part $\int_{0}^{\infty} dk$ and a left-traveling part $\int_{-\infty}^0 dk$ then we can write $V(x,t)=V_{in}(x,t)+V_{out}(x,t)$ with, for example, $V_{in}(x,t)=-i \sqrt{\frac{1}{4 \pi c}} \int_{0}^{\infty} dk \sqrt{\hbar \omega_k} \left(b_k(t)e^{i kx}-b_k^{\dagger}(t)e^{-i k x}\right)$.\\

We assume that the cavity couples to the transmission line on the right, say at $x=0$, so that incoming signals travel to the right and outgoing signals to the left. The capacitive coupling Hamiltonian between transmission line and the single mode cavity is taken to be of the form 
\begin{equation}
H_{\rm coupl}=\hbar\sqrt{\frac{\kappa v}{2\pi}} \int_{-\infty}^{\infty} dk \;(a^{\dagger} b_k+b_k^{\dagger} a),
\end{equation} 
while the Hamiltonian of the cavity field and qubit are given by $H_{\rm eff}$, Eq.~(\ref{eq:heff}). Note that $H_{\rm coupl}$ represents a simple linear coupling at $x=0$ since $H_{\rm coupl}=\hbar \sqrt{\kappa v} (a^{\dagger}b_{x=0}+b_{x=0}^{\dagger}a)$ using the spatially-labeled modes $b_x=\frac{1}{\sqrt{2\pi}} \int_{-\infty}^{\infty} dk \,e^{i kx} b_k$. \\

We can model a coherent multi-mode input state on the transmission line at initial time $t=0$ as a state $\ket{\{\alpha_k\}}=D(\{\alpha_k\}) \ket{0}$ with continuous displacement operator $D(\{\alpha_k\})=\exp(\int_{-\infty}^{\infty} dk \,(\alpha_k b_k^{\dagger}e^{-i k x}-\alpha_k^* b_ke^{i k x}))$ \cite{blow+:cont}. A plane wave with wave number $k_c$ can be modeled by taking $\alpha_k=\delta(k-k_c)(2 \pi F_l)^{1/2}$ where $F_l$ is the mean photon flux per unit length (related to the mean photon flux per unit time $F_t=vF_l$). For such a plane-wave state one has $\langle b_k^{\dagger} b_{k'} \rangle=\delta(k-k_c) \delta(k'-k_c)2 \pi F_l$ and $\langle b_x^{\dagger} b_x \rangle=F_l$.

We can also take a Gaussian pulse centered around frequency $\omega_k=|k_c| v=\omega_c$ with wavenumber $k_c > 0$ such that the pulse travels towards the cavity on the right. For such a pulse one has
\begin{equation}
\alpha_{k > 0}=\alpha \frac{e^{-(\omega_k-\omega_c)^2/W^2}}{(2\pi)^{1/4}W^{1/2}(2v)^{-1/2}}, \;\alpha_{k < 0}=0,
\label{eq:inputk}
\end{equation}
with width $W \equiv \Delta \omega \ll \omega_c$.  The coherent state $\ket{\{\alpha_k\}}=D(\{\alpha_k\}) \ket{0}$ representing this pulse will be spatially centered (with Gaussian spread) at position $x$ at time $t=0$ (due to the $x$-dependence of the displacement operator $D(\{\alpha_k\})$. The normalization of $\alpha_k$ is chosen such that the total number of photons in the pulse is $n_{\rm pulse}= \int dk \; |\alpha_k|^2=|\alpha|^2$. As the pulse travels dispersionless over the transmission line, we can drop all dependence on position $x$ or time-dependent phase-shifts when analyzing the interferometric schemes in the paper.

The relation between the cavity field and the ingoing and outgoing fields on the transmission line are usually given in terms of input and output fields $b_{in}(t)$ and $b_{out}(t)$. These operators are defined as $b_{in}(t)=-\sqrt{\frac{v}{2\pi}}\int_{-\infty}^{\infty}dk\, e^{-i\omega_k (t-t_0)} b_k(t_0)$ for $t_0 < t$ and $b_{out}(t)=\sqrt{\frac{v}{2\pi}} \int_{-\infty}^{\infty} dk\, e^{-i \omega_k(t-t_1)} b_k(t_1)$ for $t_1 > t$.  Note that as the input state at $t_0=0$ traveling towards the cavity has $\langle b_{k < 0} \rangle =0$, one could also replace the integral $\int_{-\infty}^{\infty} dk$ in $b_{in}(t)$ by $\int_{0}^{\infty} dk$ (and similarly use the integral $\int_{-\infty}^0 dk$ in $b_{out}(t))$

The Heisenberg evolution of the operators $a(t)$ is then given by (with $\tilde{\omega}_r$ defined in Eq.~(\ref{eq:defomega})):
\begin{eqnarray}
\dot{a}=-i \tilde{\omega}_r a(t)+\sqrt{\kappa} b_{in}(t)-\frac{\kappa}{2}a(t), \nonumber \\
\dot{a}=-i\tilde{\omega}_r a(t)-\sqrt{\kappa} b_{out}(t)+\frac{\kappa}{2}a(t), 
\label{eq:coupling}
\end{eqnarray}
or $b_{in}(t)+b_{out}(t)=\sqrt{\kappa} a(t)$. Thus knowing the time-dynamics of the input field $b_{in}(t)$ and the cavity field lets one determine the output field.  Conversely, knowing the output and the cavity field, one could calculate backwards to determine the dynamics of the input field. \\

In order to solve Eqs.~(\ref{eq:coupling}), one defines Fourier-transformed operators. For any time-dependent Heisenberg operator $b(t)$ one has $b[\omega]\equiv {\rm FT}(b(t))=\frac{1}{\sqrt{2\pi}} \int_{-\infty}^{\infty} dt \;e^{i \omega t} b(t)$ and $b^{\dagger}[\omega]={\rm FT}(b^{\dagger}(t))=(b[-\omega])^{\dagger}$. For a continuum of wavenumber modes $b_k$, this means that $b_k[\omega]$ has units of ${\rm sec} \times {\rm m}^{1/2}$. Note that for a discrete set of modes, such as the modes in the cavity, $a[\omega]$ has units of ${\rm sec}$ as $a(t)$ is dimensionless. For Eqs.~(\ref{eq:coupling}) one obtains
\begin{equation}
a[\omega]=\frac{\sqrt{\kappa}}{\frac{\kappa}{2}-i (\omega-\omega_r)} b_{in}[\omega] \nonumber
\end{equation}
and
\begin{equation}
b_{out}[\omega]=\frac{\kappa/2+i(\omega-\tilde{\omega}_r)}{\kappa/2-i(\omega-\tilde{\omega}_r)}b_{in}[\omega]=e^{i\varphi^{\pm}(\omega-\tilde{\omega}_{r})}b_{in}[\omega].\nonumber
\end{equation}
From the defined relations, it follows that 
\begin{equation}
b_{in}[\omega]=-\sqrt{v}\int dk \,\delta(\omega-\omega_k) b_k=-\frac{1}{\sqrt{v}} b_{k=\omega/v},\nonumber
\end{equation}
where we have restricted the input mode $b_{in}$ to only involve $k > 0$. This also gives $b_{in}^{\dagger}[\omega]=-\frac{1}{\sqrt{v}} b_{k=\omega/v}^{\dagger}$. Similarly, one has 
\begin{equation}
b_{out}[\omega]=\frac{1}{\sqrt{v}} b_{k=-\omega/v},\; b_{out}^{\dagger}[\omega]=\frac{1}{\sqrt{v}} b^{\dagger}_{k=-\omega/v}.\nonumber
\end{equation}
This shows that one may identify the input and output operators with mode operators corresponding to a certain wave number and thus with a certain frequency $\omega$. It also allows us to translate the input state in Eq.~(\ref{eq:inputk}) into expectations of $b_{in}[\omega]$ etc. Note that $b_{in}[\omega]$ (and $b_{out}[\omega]$) have units of ${\rm sec}^{1/2}$ and obey $[b_{in}({\omega}), b_{in}^{\dagger}({\omega'})]=\delta(\omega-\omega')$ via the commutation relations for the mode operators $b_k$.

The mode transformations of the (degenerate) parametric amplifiers in Eqs.~(\ref{eq:pa}, \ref{eq:dpa}) are given in terms of the operators $a_i(\omega)$, but these operators should be similarly interpreted as input-output operators $a_{i,in}[\omega]$ and $a_{i,out}[\omega]$ (see e.g. \cite{YB:amp}). Thus in the main text we simply refer to such continuum-mode operators as $a_i(\omega), b_{in}(\omega)$ (with round brackets) etc. with expectations $\langle b_{in}(\omega)\rangle=\alpha(\omega)$ and units of ${\rm sec}^{1/2}$.


\begin{thebibliography}{43}
\bibitem{foot1}The usual Mach-Zehnder interferometer is called a SU(2) interferometer as the Yurke {\em et al.} analysis \cite{YMK} shows that the transformation induced by the interferometer is a rotation on angular momenta observables $(J_x,J_y,J_z)$ which characterize the input to the interferometer. For the SU(1,1) interferometer, the induced transformation on the input observables is that of the Lorentz group SU(1,1).
\bibitem{foot3}The derivation of the stochastic master equation describing a homodyne measurement, see  \cite{book:wm}, in which the Gaussian fluctuations on the outgoing quadrature are due to the Gaussian fluctuations around the photon number of the local oscillator, is thus also, strictly speaking, not applicable here. It is rather the sequence of amplifiers that should be described within such a stochastic master equation formalism.
\end{thebibliography}
\end{document}